\documentclass[a4paper]{article}

\usepackage{lineno,hyperref,amsmath,graphics,dcolumn}
\usepackage{tabularx}
\usepackage{array}
\usepackage{rotating} 
\usepackage{booktabs} 
\usepackage{multirow} 
\usepackage{subfig}
\usepackage{bm}
\usepackage{amsmath} 

\usepackage{changes}
\usepackage{natbib}
\graphicspath{{./grafici/}}
\modulolinenumbers[1]

\title{A spatio-temporal analysis of NO$_{2}$ concentrations during the Italian 2020 COVID-19 lockdown}
\author{Guido Fioravanti$^1$ \thanks{Corresponding author Guido Fioravanti: guido.fioravanti@isprambiente.it} \and Michela Cameletti$^2$ \and Sara Martino$^3$ \and Giorgio Cattani$^1$ \and Enrico Pisoni$^4$}
\date{%
    $^1$ Italian National Institute for Environmental Protection and Research, Department for environmental evaluation, control and sustainability. Rome, Italy \\%
    $^2$ Department of Economics, University of Bergamo, Bergamo, Italy\\
    $^3$ Norwegian University of Science and Technology, Trondheim, Norway\\
    $^4$ European Commission, Joint Research Centre, Ispra, Italy\\[2ex]%
    \today
}

\begin{document}

%
%
%
%

\maketitle

\abstract{When a new environmental policy or a specific intervention is taken in order to improve air quality, it is paramount to assess and quantify - in space and time - the effectiveness of the adopted strategy. The lockdown measures taken worldwide in 2020 to reduce the spread of the SARS-CoV-2 virus can be envisioned as a policy intervention with an indirect effect on air quality. In this paper we propose a statistical spatio-temporal model as a tool for intervention analysis, able to take into account the effect of weather and other confounding factors, as well as the spatial and temporal correlation existing in the data. In particular, we focus here on the 2019/2020 relative change in nitrogen dioxide (NO$_{2}$) concentrations in the north of Italy, for the period of March and April during which the lockdown measure was in force. As an output, we provide a collection of weekly continuous maps, describing the spatial pattern of the NO$_2$ 2019/2020 relative changes. We found that during March and April 2020 most of the studied area is characterized by negative relative changes (median values around -25\%), with the exception of the first week of March and the fourth week of April (median values around  5\%). As these changes cannot be attributed to a weather effect, it is likely that they are a byproduct of the lockdown measures.}



\maketitle

\section{Introduction}

In the last decades a large number of initiatives have been taken in European countries in order to reduce air pollution and its adverse effects on human health. This is for example the case of low emission zones or congestion charge introduced in urban areas \citep[see e.g.,][]{Fasso2013,HOLMAN2015, Maranzano2020}, of more stringent limits on the content of sulfur in marine fuels \citep[e.g.,][]{GRANGE2019} or of new air pollution control
regulations \citep[e.g.,][]{FONT2016, HAN2021}, among others \citep[see][for a review]{BURNS2020}. A crucial issue in air quality management is the assessment of the efficacy of a specific intervention or a new environmental policy. In particular, quantifying the effective changes in pollutant concentrations due to the adopted strategy can be difficult because of the complexity of air pollution dynamics, strongly depending on weather conditions. Moreover, when the intervention is time-delimited and characterizes different areas, it may be interesting to evaluate the variability in space and time of its effect, if any.

The lockdown measures adopted by many countries in 2020 in order to to prevent the spread of the SARS-CoV-2 virus can be considered as a policy intervention, with a possible indirect effect on air quality. In this regard, the main focus is on assessing how the lockdown  affected air pollution levels, in particular after controlling for meteorology, long term trend and other confounding factors. The literature on this topic is obviously quite recent but already large, as discussed in the two recent systematic reviews by \cite{Gkatzelis2021} and \cite{Rana2021}.

In this paper, we propose a new statistical modeling approach for assessing and quantifying the effectiveness of a policy intervention on air quality. In particular, our statistical model is defined for daily differences of pollutant concentrations and has a spatio-temporal specification which gives us the possibility to estimate in space and time the relative change in air pollution levels. We show here an application of our modelling strategy for nitrogen dioxide (NO$_{2}$) concentrations in northern Italy during the months of the first Italian lockdown (March and April 2020) as compared to 2019. In early 2020, Italy was the first European country to adopt strict lockdown measures \citep{Remuzzi2020}. Starting from late February 2020, people were banned from travelling and all the non essential socio-economic activities were stopped. These restrictive measures were initially adopted in some limited municipalities in northern Italy and then were extended to the entire country with the national lockdown in force since March 8$^{\text{th}}$, 2020 \citep{Malpede2020,Sanfelici2020}. The restrictions caused, incidentally, a strong decrease in anthropogenic emissions of the main air pollutants, especially for some sectors such as road transport and aviation.

 Several environmental studies revealed that restrictions on mobility during the lockdown had a positive effect on NO$_2$ levels, even if not homogeneously across the considered spatial domains. For example, the broadly-publicized data from the Copernicus Sentinel-5P satellite \citep{Dutta2021, Muhammad2020,Bar2021} recorded for NO$_{2}$  a sharp drop,  in the range 20–55\%, during January – April 2020 compared to 2019 in many cities in China, India, Western Europe and United States. With regards to Italy, \citet{Bauwens2020} found the average TROPOMI NO$_{2}$ column during the lockdown period in 2020 to be between 38 ($\pm$ 10\%) lower than during the same period in 2019 in Milan. Similarly, \cite{Cersosimo2020} reported a general decrease in NO$_{2}$ levels over the Po Valley during the lockdown using both satellite and in situ measurements. For the city of Rome and its surroundings, \citet{Bassani2021} documented a 2019 vs 2020 reduction of 50\% in NO$_{2}$ concentrations using surface measurements from urban traffic stations. The NO$_{2}$ decrease for the city of Rome was also documented in \cite{Kumari2020}. Finally, a general decrease in the NO$_{2}$ levels for three cities in Tuscany region was described by \cite{Donzelli2020}, for Reggio Emilia by
\cite{Marinello2021} and for Naples by \cite{Sannino2021}. The main weakness of the aforementioned studies consists in their descriptive approach, based on the direct comparison of pollutant concentrations before and after the lockdown or between the lockdown period and the corresponding period of the previous year(s). In other words, these studies make no attempt to adjust for the effect of meteorological conditions which can be adverse or favorable to pollutant dispersion. In this regard, it is worth to note that the first quarter of 2020 experienced exceptional weather conditions with also stronger positive temperatures anomalies over Europe \citep{Barre2020,von2021}.

A number of studies approached the problem of assessing the lockdown effects on air quality using chemical transport models (CTM) in order to compare  model forecasts under the business-as-usual (BAU) emission scenario with the observed ground-level measurements or with the expected concentrations computed under a COVID-19 pandemic scenario with reduced emissions \citep[see e.g.,][]{Barre2020, Menut2020, Piccoli2020, Putaud2020,WANG2021}. In any case, the use of deterministic models poses a number of practical and conceptual difficulties. First, collecting the needed input data (e.g., emission inventories and meteorology data) is far from straightforward. Second, deterministic models are complex to be run and are not able to properly assess the uncertainty of the results.

Another substantial research line is represented by model-based studies. In this context historical measurement data from previous years (or from the pre-lockdown period) are used to run machine learning algorithms \citep[see e.g.,][]{Barre2020, Diemoz2021, Keller2021, KIM2021, Grange2021, Petetin2020, Granella2021} or to estimate statistical models,  as multiple linear regression models \citep[e.g.,][]{BAO2020, Dacre2020, HOERMANN2021}, Generalized Additive Models \citep[e.g.,][]{EEA:2020, ORDONEZ2020, Solberg2021, Sverre2021} or Autoregressive Integrated Moving Average models \citep[e.g.,][]{Tyagi2020}. The fitted model is then used to predict concentrations for 2020 (or for the post-lockdown period) under the BAU (or counterfactual) scenario, i.e., assuming that the lockdown did not take place. In order to correct for the weather effect and temporal trends, meteorological and time variables are included in the model as linear or non-linear effects. The differences between the predictions, derived from the estimated model, and the observed concentrations (i.e., the out-of-sample prediction errors) are then used to evaluate the effect of the lockdown restrictions. Other papers adopt a different modeling approach and define the counterfactual by using data from cities not subject to lockdown restrictions; then difference-in-differences (DID) models are used to estimate the relative change in pollutant concentrations in the treatment group (locked-down cities) compared with the control group (non-locked-down cities) \citep[see e.g.,][]{Guojun2020, WangM2021}. An extension of the DID method is proposed in \cite{Zheng2021} with the aim of computing the would-be average concentrations without the COVID-19 pandemic by removing the meteorological confounding and accounting for the temporal trend.
Another modeling strategy makes use of time series model or dynamic panel data model, where pollutant concentrations, measured in 2020 and possibly in previous years, represent the response variable. The lockdown effect is included as a time-dependent dummy variable in the set of regressors, together with meteorological and time variables \citep[see e.g.,][]{BAO2020, BELOCONI2021, Cameletti2020}. In this case the effectiveness of the lockdown is evaluated by means of the lockdown dummy coefficient and its interaction with time or other regressors.

Whatever the adopted model-based strategy is, pollutant concentrations time series can be analysed separately for each monitoring station or jointly. The second solution leads to more efficient parameter estimates and a better predictive capability because of the larger amount of available data. More importantly, when measurements come from several spatial sites it is convenient  to account also for spatial correlation, besides the temporal one, for explaining any residual variability. This is a standard and well-established option in air pollution modeling  \citep[see e.g.,][]{Finazzi2013, sahu2006, Lee2016} given that nearby monitoring stations are expected to show similar pollutant concentrations values. As far as we know, \cite{BELOCONI2021} is the only study which implements a spatio-temporal model for evaluating the effect of COVID-19 lockdown on air quality, while all the other papers fail to consider the spatial correlation. However, in \cite{BELOCONI2021} the analysis of the impact of the lockdown is limited at the single point stations and no continuous maps are provided for the entire region of interest.

In this paper, using a spatio-temporal statistical model, we aim to produce spatio-temporal maps showing continuously in space and across time the impact of the lockdown on air pollution. We expect these highly spatially resolved maps to help in assessing if the lockdown effect was homogeneous in the considered area or was more consistent in particular zones. Having this goal in mind, we modeled the 2019/2020 daily differences of NO$_2$ concentrations (in March and April), rather than jointly modeling the data available for the two years. This makes it possible to focus on the change occurred between 2019 and 2020 in the NO$_2$ levels, while still accounting for the effect of meteorology. In particular, by including a spatial stochastic component, our model is able to take care of the spatial correlation between observations and generate spatially continuous prediction surfaces, also where no ground-monitoring stations are available and in remote or mountainous areas \citep{Diemoz2021}. The model we propose is applied here to the Italian COVID-19 case study but it is a general modeling approach that can be implemented anytime it is necessary to evaluate the effect of an intervention on an output variable with a spatio-temporal dimension.

The paper is structured as follows: in Section \ref{Data} we introduce the study domain and the input data. Then, in Section \ref{Sec:model} the proposed spatio-temporal regression model is described; details are also provided with respect to the prediction phase of the analysis. Finally, in Section \ref{Sec:results} we describe the main results of our analysis with particular attention to the prediction maps. Section \ref{Sec:Conclusions} concludes the paper.


\section{Data material}\label{Data}

\subsection{Monitoring sites and air pollution data}

In this study we focused on the Italian COVID-19 lockdown period (from  March $1^{\text{st}}$ to 30 April $30^{\text{th}}$, 2020), corresponding to 10 weeks of daily observations. The 2019 and 2020 raw NO$_2$ hourly data (expressed as $\mu g/m^3$), together with monitoring stations metadata, were extracted from the national database used to store and process the Italian Air quality measurements \citep{SNPA2020}. 
The hourly data were measured at stations operated by the local Regional Environmental Protection Agencies, following the European standards EN 14211:2012 for NO, NO$_2$, and NO$_\text{X}$ \citep{CEN-TC264}.
All ground concentrations were fully validated following internal quality assurance and quality control standard procedures. 

\begin{figure}
\includegraphics[width=\columnwidth]{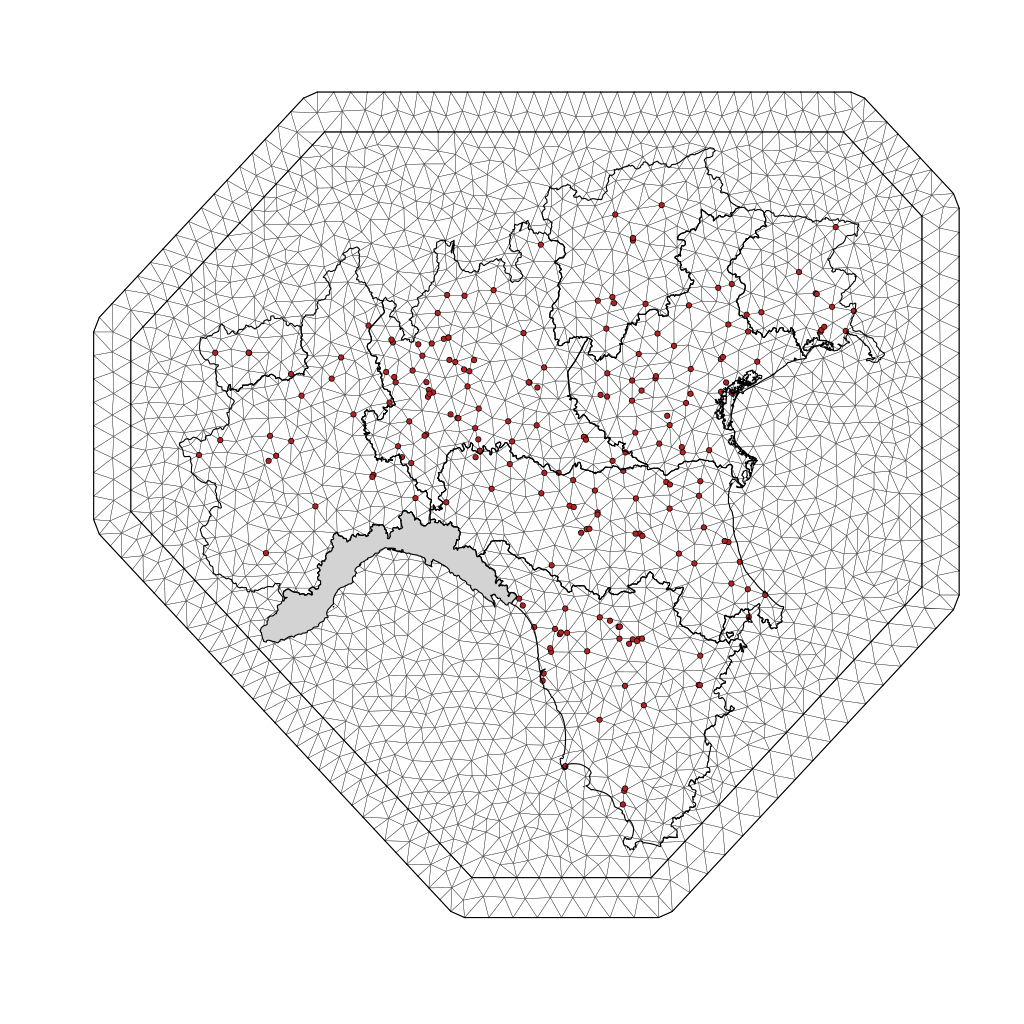}
\caption{Spatial distribution of the monitoring sites for $\text{NO}_{2}$ concentrations (red points) and mesh adopted for the implementation of the spatio-temporal model described in Section \ref{Sec:model}.}\label{fig:mesh}
\end{figure}

Daily averages were computed provided that a daily 75\% data coverages was achieved (i.e., at least 18 valid hourly records out of 24). The input dataset for our analysis regards 200 monitoring sites with a low proportion of missing data ($< 25\%$ per station) and distributed across 8 regions or autonomous provinces in the north of Italy (Valle d'Aosta 4 stations, Piedmont 17, Veneto 32, Lombardy 55, Autonomous province of Trento 5, Friuli Venezia Giulia 12, Autonomous province of Bozen 5, Emilia Romagna 36) plus Tuscany region in the centre of Italy (with 34 stations). The spatial distribution of the selected  stations is illustrated in Figure~\ref{fig:mesh}. The monitoring stations cover urban (123 stations), suburban (39) and rural (38) areas. At the time of the analysis, we were not able to access to the NO$_{2}$ data from Liguria region (the western region filled in gray in the map in Figure~\ref{fig:mesh}).

The investigated area is characterized: in the north, by remote and relatively pristine areas with the high peaks of the Alps mountain system; in the southern part, by the mountains and hills of the Apennines range (the "spine" of the Italian peninsula); in the centre, by the Po Valley. The latter is a large area which includes Lombardy,  Emilia-Romagna,  Veneto  and  Piedmont, regions which account for 48.2\% of Italy’s GDP \citep{OECD2020}. Interestingly, in early March 2020, these were the regions with the highest incidence of COVID-19 cases\footnote{\url{https://www.salute.gov.it/imgs/C_17_pagineAree_5351_0_file.pdf}}. The Po Valley exhibits intense human pressure and severe pollution levels, with high population density, urban sprawl, intensive agricultural practice and livestock management \citep{Pezzagno2020, Romano2020}. This leads to large amounts of NO$_{\text{X}}$ emissions from vehicles, methane and ammonia from agricultural activities and particulate matter (PM) from residential heating. The Alps and the Apennines often limit the air flows between the Po valley and the rest of continental Europe causing air pollution stagnation. 

Table~\ref{tab:summary_no2} provides the summary statistics for the 2019 and 2020 NO$_{2}$ daily concentrations. A certain variability in the NO$_{2}$ values is observed  across areas, month and year. Nonetheless, it is apparent that lower NO$_{2}$ measurements characterize the 2020 months compared with 2019. For example, the monthly NO$_{2}$ mean values are in the range 13 - 35 $\mu g/m^3$ in March 2019 and  12 - 26 $\mu g/m^3$ in April 2019, while in March and April 2020 the same values vary between 11 - 24 $\mu g/m^3$ and 7 - 17 $\mu g/m^3$, respectively. This means that for the lockdown months of 2020 the range of the monthly mean concentrations is approximately halved compared with the same months of the previous year. The same situation can be observed also in the median and maximum values.

\begin{sidewaystable}\footnotesize
\begin{center}
\begin{tabular}{clccccccccccc}   
\toprule
\multicolumn{1}{c}{\phantom{Ciao}} &  \multicolumn{1}{c}{\textbf{Area}} & \multicolumn{5}{c}{\textbf{2019}} & \multicolumn{1}{c}{\phantom{abc}} & \multicolumn{5}{c}{\textbf{2020}}\\
\cmidrule{1-2} \cmidrule{3-7} \cmidrule{9-13}\\
\phantom{Month} & \phantom{Region} & Mean & SD & Min & Median & Max & \phantom{abc} & Mean & SD & Min & Median & Max \\ 
March & Bolzano (AP) & 34.7 & 11.5 & 11 & 33 & 72 & \phantom{abc} & 24.2 & 11.6 & 4 & 22 & 55 \\ 
\phantom{March} & Trento (AP) & 30.7 & 17.8 & 3 & 30 & 92 & \phantom{abc}  & 19.6 & 12.5 & 3 & 18 & 61 \\ 
\phantom{March} & Emilia Romagna & 24.8 & 16.0 & <1 & 22 & 85 &  \phantom{abc} & 16.3 & 10.6 & <1 & 14 & 57 \\ 
\phantom{March} & Friuli Venezia Giulia & 21.1 & 9.9 & 3 & 20 & 53 &  \phantom{abc} & 14.2 & 9.1 & <1 & 13 & 42 \\ 
\phantom{March} & Lombardy & 32.1 & 15.5 & 2 & 30 & 99 &  \phantom{abc} & 22 & 11.6 & 2 & 20 & 85 \\ 
\phantom{March} & Piedmont & 27.5 & 16.3 & 3 & 24 & 92 & \phantom{abc} & 17.8 & 9.1 & 3 & 16 & 45 \\ 
\phantom{March} & Tuscany & 21.6 & 13.9 & <1 & 20 & 80 & \phantom{abc} & 15.5 & 10.7 & <1 & 14 & 62 \\ 
\phantom{March} & Valle d'Aosta & 13.0 & 8.5 & 1 & 11.5 & 33 & \phantom{abc} & 11.1 & 9.1 & 1 & 10.0 & 50 \\ 
\phantom{March} & Veneto & 27.1 & 12.6 & 3 & 27 & 68 &  \phantom{abc} & 17.8 & 10.1 & 2 & 16 & 49 \\ 
April & Bolzano (AP) & 26.2 & 9.2 & 7 & 25 & 50 & \phantom{abc} & 16.8 & 4.8 & 6 & 17 & 29 \\ 
\phantom{April} & Trento (AP) & 22.7 & 13.7 & 3 & 21.5 & 58 &  \phantom{abc} & 13.5 & 6.2 & 4 & 13 & 33 \\ 
\phantom{April} & Emilia Romagna & 19.3 & 13.2 & <1 & 16 & 72 &  \phantom{abc} & 10.9 & 6.6 & <1 & 9 & 40 \\ 
\phantom{April} & Friuli Venezia Giulia & 14.2 & 7.8 & 1 & 13 & 37 &  \phantom{abc} & 9.7 & 4 & 1 & 9 & 25 \\ 
\phantom{April} & Lombardy & 23.0 & 13.6 & 1 & 20 & 97 &  \phantom{abc} & 15.6 & 7.8 & 1 & 14 & 58 \\ 
\phantom{April} & Piedmont & 19.8 & 10 & 3 & 18 & 73 &  \phantom{abc} & 11.9 & 6.1 & 1 & 11 & 31 \\ 
\phantom{April} & Tuscany & 19.7 & 13.3 & <1 & 17 & 83 &  \phantom{abc} & 10.1 & 6.2 & 1 & 9 & 33 \\ 
\phantom{April} & Valle d'Aosta & 12.0 & 7.9 & <1 & 11 & 32 &  \phantom{abc} & 6.8 & 3.5 & 1 & 6 & 14 \\ 
\phantom{April} & Veneto & 18.3 & 11.2 & 2 & 16 & 62 &  \phantom{abc} & 11.6 & 6.3 & 2 & 10 & 35 \\ 
\hline \\[-1.8ex] 
\end{tabular}
\caption{Summary statistics - mean, standard deviation (SD), minimum (Min), median, and maximum (Max) - for NO$_{2}$ concentrations (in $\mu g/m^3$), for the considered region or autonomous province (AP) and for March and April of years 2019 and 2020.} 
\label{tab:summary_no2} 
\end{center}
\end{sidewaystable} 

\subsubsection{Daily differences}\label{Sec:dailydifferences}

To match the 2019 and 2020 daily NO$_2$ concentrations and compute the corresponding daily differences, we excluded the use of the calendar date. The reason for this choice is that we don't want to match business days in 2020 with weekend days in 2019, and vice versa, given the well-known NO$_2$ weekly cycle (characterized by lower values during the weekend). Rather, we aligned the 2019  daily measurements of each monitoring site by week number (according to the ISO-8601 standard) and day of the week (Monday, Tuesday, \dots, Sunday) taking as a reference those of the 2020 daily measurements. A similar approach is described in \cite{Ruan2020} to analyze the impact of COVID-19 on electricity consumption in the US. By doing so, we were able to compare the first Monday of the first week in March 2020 with the first Monday of the first week in March 2019, the first Tuesday of the first week in March 2020 with the first Tuesday of the first week in March 2019, and so on. However,  even using this approach it can happen that a public national holiday (e.g., Easter Monday or April 25$^{\text{th}}$) in 2020 is matched with a working day in 2019, and vice versa. 

\begin{figure}[!ht]
\includegraphics[width=\textwidth]{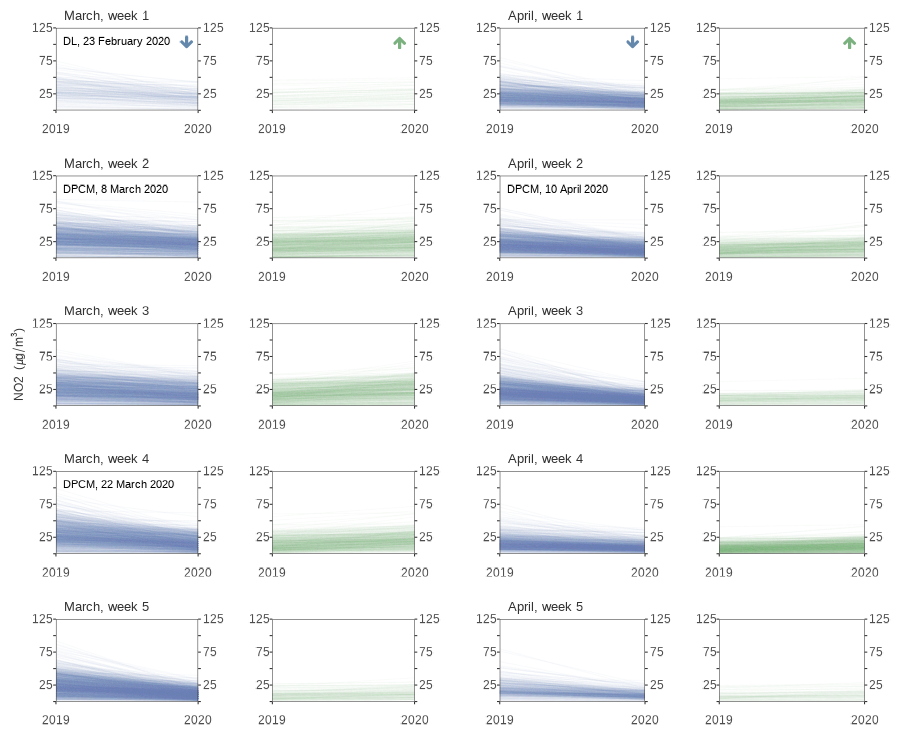}
\caption{Aligned 2019-2020 daily NO$_{2}$ concentrations during the weeks of March and April. Each line represents a couple of aligned data of a monitoring site, distinguishing between decrements (blue lines) and increments (green lines) in NO$_{2}$ levels. The different government decrees and rules of the Italian lockdown are reported as text in the boxes. Each weekly plot depicts 1400 lines (200 monitoring sites times 7 days). Note that after the alignment process the first week of March contains just one day (Sunday, 200 lines) and the last week of April only two days (Monday and Thursday, 400 lines). }\label{fig:serie_osservate}
\end{figure}

The parallel plot in Figure \ref{fig:serie_osservate} graphically shows the 2019 and 2020 NO$_{2}$ aligned datasets. Here, the daily data are visually aggregated at the weekly level. The plot nicely reveals the decrease in the NO$_2$ concentrations during March and April 2020, compared with 2019, especially during the last two weeks of March. At the same time, the  positive increments occurring across the weeks in some of the stations suggest that it would be wrong to think that, if the lockdown affected the NO$_{2}$ concentrations levels, such an effect was a homogeneous phenomenon both in time and space.

We conclude this section observing that a small fraction (about $3.5\%$) of aligned daily measurements exhibits unusual combinations of extremely high values in one year and extremely low values in the other. These combinations correspond to 2019/2020 relative changes greater than 100\% in absolute value, which abnormally lie outside the general pattern seen in our input dataset. These data have no meaning for the purposes of our analysis and were discarded.

\subsection{Spatial and spatio-temporal regressors}\label{Sec:regressors}

For each monitoring site, we retrieved its meteorological conditions and spatial characteristics for a total number of 13 independent explanatory variables (see Table~\ref{table:predittori}), whose selection was driven by expert knowledge \citep{FIORAVANTI2021}. 

To describe part of the spatial variation of the NO$_{2}$ daily differences, three  geographical variables (constant in time) were considered: the linear distance from the the nearest major road, the altitude and the percentage of agricultural and arable lands. \cite{YEGANEH2018} found that traffic volume and congestion data for all of the individual roads can effectively improve the spatio-temporal modelling of NO$_{2}$ concentrations. Unfortunately, this information is not available for the whole investigated domain. 

How weather affects NO$_{2}$ concentrations is assessed using 10 variables (such as total precipitation, wind speed, relative humidity, surface pressure) retrieved from the ERA5-Land dataset from the  Copernicus Climate Data Store (\url{https://climate.copernicus.eu/}). ERA5-Land is a regridded version ($\sim$9 Km grid spacing) of the ERA5 climate reanalysis \citep{Hersbach_et_al_2020} of the European Centre for Medium-Range Weather Forecasts (ECMWF). The use of ERA5 data is documented, among others, by \cite{Chan2021} for investigating the effect of COVID-19 pandemic on NO$_{2}$ concentrations over Germany. 

At this point it is important to stress that the meteorological variables enter the model as $2020-2019$ daily differences. More precisely, for the meteorological daily data we applied the same alignment procedure used for the NO$_{2}$ concentrations and described in Section~\ref{Sec:dailydifferences}. Afterwards, we calculated the daily differences which we used as model regressors, as explained in Section~\ref{Sec:model}, where the details of the statistical model are given. It is worth to note that the inclusion in the model of the differenced meteorological variables alleviates potential multicollinearity problems and makes it possible to include in the same model regressors like temperature and surface pressure, which otherwise would be highly correlated.

\begin{table}  
\begin{tabular}{@{\extracolsep{0.5pt}} lllll@{} }
   \toprule
   Data Source & \phantom{abc} & Description & Unit & Spatial Resolution\\
   \midrule
   \multicolumn{5}{l}{\textbf{ERA5}}\\
    \multicolumn{5}{c}{\phantom{abc}}\\
    \phantom{a} && Minimum Planet Boundary Layer & Km & \multirow{11}{*}{$\sim9$ Km}\\ 
    \phantom{a} && Maximum Planet Boundary Layer  & Km & \\ 
    \phantom{a} && Total Precipitation & mm & \\        
    \phantom{a} && Surface Pressure & hPa & \\
    \phantom{a} && Average Temperature at 2 meters & $^{\circ}$C & \\
    \phantom{a} && Wind Speed & m/s & \\
    \phantom{a} && Previous day Wind Speed & m/s & \\
    \phantom{a} && Relative Humidity & \% & \\
    \phantom{a} && Net Irradiance & $W/m^{2}$ & \\
    \phantom{a} && Diurnal Temperature Range & $^{\circ}$C & \\
    \multicolumn{5}{c}{\phantom{abc}}\\
    \multicolumn{5}{l}{\textbf{Global Multi-resolution Terrain Elevation Data}}\\
    \multicolumn{5}{c}{\phantom{abc}}\\   
    \phantom{a} && Altitude & m &  ~1 Km\\
    \multicolumn{5}{c}{\phantom{abc}}\\
    \multicolumn{5}{l}{\textbf{OpenStreetMap}}\\
    \multicolumn{5}{c}{\phantom{abc}}\\
    \phantom{a} && Linear distance to the nearest primary road & m & 1 km\\
    \multicolumn{5}{l}{\textbf{Corine Land Cover}}\\
    \multicolumn{5}{c}{\phantom{abc}}\\
    \phantom{a} && Agricultural and arable lands & \% & 1 km\\
    \bottomrule
\end{tabular}
 \caption{Spatial and spatio-temporal predictors included in the model.}\label{table:predittori}
\end{table}


\section{Bayesian spatio-temporal model}\label{Sec:model}

Consider a couple of NO$_2$ concentrations $y^m_{2019}(t,s_i)$ and $y^m_{2020}(t,s_i)$ temporally aligned according to the procedure described in Section \ref{Sec:dailydifferences}, where $s_i$ denotes the location (with $i=1,\ldots,200$) and $t$ the day ($t=1,\ldots, T^m$) of month $m=1,2$ (where $m=1$ denotes March) of year 2019 and 2020, respectively. 

As the objective of this study is to evaluate the change in  NO$_2$ levels between 2019 and 2020, we first of all defined the daily differences of the log-transformed NO$_2$ concentrations:

\begin{equation}
\Delta^m(t,s_i)=\log\left(y^m_{2020}(t,s_i)\right) - \log\left(y^m_{2019}(t,s_i)\right)
=\log\left(\frac{y^m_{2020}(t,s_i)}{y^m_{2019}(t,s_i)}\right).\label{eq:modelEQ1}
\end{equation}

The logarithmic transformation is a common choice in air pollution analysis in order to reduce the typical positive skewness observed in concentrations distributions \citep{OTT1990}.

As the Italian lockdown fell almost at the end of the winter season, when usually the meteorological conditions favor the dispersion of the pollutants, a downward trend in NO$_{2}$ concentrations is expected across March and April. This could have a confounding effect when trying to isolate the impact of COVID-19 lockdown measures on air quality. In order to control for this long-term trend, we decided to model the daily differences separately for the two months (March and April).

The measurement equation is given by
\begin{equation}\label{eq:modelEQ2}
\Delta(t,s_i)  = \mu(t,s_i) + \epsilon(t,s_i),
\end{equation}
where $\epsilon(t,s_i)\sim N(0,\sigma^2_\epsilon)$ is the Gaussian measurement error assumed to be independent in space and time. For the sake of simplicity in Equation~\eqref{eq:modelEQ2}  we omitted the index $m$ given that the model structure is the same for the considered months. 

The mean level $\mu(t,s_i)$ is then defined as the sum of fixed and random effects as follows:
\begin{equation}\label{eq:modelEQ3}
\mu(t,s_i)  = \alpha_0+ \alpha_1 t+ \gamma {I}^S_t+ \bm{z}(s_i)\boldsymbol{\beta}^\prime_z  +\bm{\Delta}_ x(t,s_i)\boldsymbol{\beta}^\prime_x+ v(s_i) + u(t,s_i),
\end{equation}
where $\alpha_0$ is the intercept and $\alpha_1 t$ the linear temporal trend which should account for the short term variation across days in the month. The term  
$\bm{z}(s_i)$ is the $p_ z$-dimensional vector of the purely spatial regressors, while $\bm{\Delta}_ x(s_i,t)$ is the $p_x$-dimensional vector of the differences computed using the values of the meteorological regressors, as described in Section \ref{Sec:regressors}. We assume a linear effect for the spatial and spatio-temporal covariates by means of the parameters' vector $\bm{\beta}_z$ and $\bm \beta_x$, respectively. 
In order to test whether or not there was a variation in the Sunday effect  during the lockdown restrictions, we included also the dummy variable ${I}_t^S$ which is equal to 1 when $t$ is Sunday and zero otherwise. The corresponding coefficient $\gamma$ represents the additional expected change in the mean difference $\mu(t,s_i)$ during Sunday.
The term $v(s_i)\sim N(0,\sigma^2_v)$ represents a temporally and spatially uncorrelated Gaussian random effect which captures some of the small scale spatial variability. 
Finally, $u(t, s_i)$ is the residual space-time correlation for which the following first order autoregressive dynamics was specified:
\[u(t,s_i)=a\; u(t-1,s_i)+\omega(t,s_i),\]
for $t = 2,\ldots, T$ and $|a| < 1$.
The innovations $\omega(t,s_i)$ have a zero-mean Gaussian distribution, are uncorrelated in time (i.e., $\text{Cov}\left(\omega(t,s_i),\omega(t^\prime,s_i)\right)=0$ if $t\neq t^\prime$) while being spatially correlated, i.e., $\text{Cov}\left(\omega(t,s_i),\omega(t,s_j)\right) = \mathcal{C}(h)$, where $h$ is the Euclidean distance between site $i$ and $j$ and $\mathcal{C}(h)$ is the Mat\'ern covariance function with variance $\sigma^2_\omega$ and range $\rho$. For more details about this separable spatio-temporal covariance structure see for example  \cite{Cameletti:2011,Cameletti2013}.

\subsection{Prior specification and model implementation}

The model described in Section \ref{Sec:model} is developed in the Bayesian framework and is fully specified once prior distributions are set. 

Vague Gaussian  priors were used for $\alpha_0$, $\alpha_1$ and for the elements of $\bm{\beta}_z$ and $\bm{\beta}_x$. For the remaining parameters Penalized Complexity (PC) priors \citep{simpson2017} were used. PC priors are a relatively new approach to specify weakly informative priors \citep{Lemoine2019} in realistically complex statistical models with the twofold purpose of penalizing model complexity and avoiding overfitting. 

For the standard deviation parameters (here $\sigma_\epsilon$ and $\sigma_v$) PC priors are generally defined as $\text{Prob}(\sigma>u_\sigma) = \alpha_\sigma$, where $u_\sigma>0$ is a quantile of the prior and $0\leq\alpha_\sigma\leq 1$ is a probability value. In our analysis we set $u_\sigma = 1$ and $\alpha_\sigma = 0.1$ for both $\sigma_\epsilon$ and $\sigma_v$. This choice can be motivated considering that the total standard deviation of the observed daily differences of the log-transformed NO$_{2}$ concentrations is $\sim0.5$ in each month, so it is very likely that the  variance of each component is lower than 1.

The joint PC prior suggested in \citet{Fuglstad2019} was used for $\rho$ and $\sigma_\omega$. This can be specified through
\[\text{Prob}(\rho < u_\rho) = \alpha_\rho;\text{  } \text{Prob}(\sigma_\omega > u_{\sigma_\omega}) = \alpha_{\sigma_\omega},\] 
where we set $u_\rho = 150$, $\alpha_\rho = 0.8$, $u_{\sigma_\omega} = 1$, $\alpha_{\sigma_\omega} = 0.01$. Finally, for the autocorrelation parameter $a$ we used the PC prior proposed in \citet{Sorbye_ar1}. This can be specified through $\text{Prob}(a>u_a)=\alpha_a$, where we set $u_a = 0.8$ and $\alpha_a = 0.3$. All these choices reflect both previous findings \citep[see e.g.][]{Cameletti2013, FIORAVANTI2021} and restrictions to the possible values of $u_a$ and $\alpha_a$.

Inference was carried out by using the INLA-SPDE approach \citep{Martino2020, Lindgren2015}, which has been proved to be computationally faster than the Markov chain Monte Carlo (MCMC) approach commonly used for Bayesian inference.

\subsection{Prediction and posterior summary statistics}\label{Sec:postsummary}

Once the model has been fitted to the observed data, we used Monte Carlo (MC) simulation to generate a large number of samples (say 1000) from the posterior predictive distribution of $\Delta(t,s)$ for any location $s$ in the study area and day $t$:
\begin{equation}\label{eq:postpreddistr}
p\left(\Delta(t,s\right)\mid \mathcal D) = \int p(\Delta(t,s)\mid \bm{\theta})p(\bm{\theta}\mid \mathcal{D}),
\end{equation}
where $\bm \theta=\left(\alpha_0,\alpha_1,\gamma, \bm\beta_x, \bm\beta_z, a,\rho, u(t,s), \sigma^2_\epsilon, \sigma^2_v, \sigma^2_\omega\right)$ is the vector of all the model parameters, whose
uncertainty  is averaged out given all the observed data 
$\mathcal{D}$. For simplicity the generic sampled value will be denoted by $\hat \Delta(t,s)$. 

To generate prediction maps, we used a raster grid with a spatial resolution of 1 km. For each pixel of this grid, corresponding to the generic location $s$, we simulated 1000 values $\hat \Delta(t,s)$ from Equation~\eqref{eq:postpreddistr}. While computing the predictions, we set the values of $\bm{\Delta}_ x(s,t)$ equal to zero for each $s$ and $t$. This corresponds to assume that the meteorological conditions are equal in 2019 and 2020 for each location and time point. 

We derived the posterior distribution of the relative change of NO$_2$ concentrations between 2019 and 2020 by using the following transformation:
\begin{equation}\label{eq:exptrans}
\tilde\Delta(t,s)=\exp\left(\hat\Delta(t,s)\right)-1=\frac{\hat y_{2020}(t,s)}{\hat y_{2019}(t,s)}-1,
\end{equation}
where $\tilde\Delta(t,s)$ takes negative (positive) values if lower (higher) NO$_2$ concentrations are expected in 2020 compared to 2019, and equal to zero in case of no change.

Finally, we averaged the 1000 MC samples from the daily distributions of $\tilde\Delta(t,s)$ at the week temporal resolution. The result is a collection of 1000 predicted raster maps for each week, from which it was possible to obtain the maps of the posterior mean and of the 2.5\% and 97.5\% quantiles. These latter two identify the bulk of the posterior distribution of each grid cell and were used to determine the 95\% credible interval. When a credible interval does not include the zero, the corresponding relative change statistically differs from zero at the significance level of 0.05.

Finally, it is noteworthy to say that to take into account the weekly cycle of NO$_{2}$ concentrations, we generated both mean and quantile maps distinguishing between working days (Monday - Saturday) and weekends (Sunday).

\subsection{Implementation and data availability}

R software was used for the model implementation by means of the \texttt{R-INLA} package (\url{https://www.r-inla.org}). For the manipulation of the raster maps, we used the R \texttt{raster} package (\url{https://cran.r-project.org/web/packages/raster/index.html}) and the \texttt{CDO} software (\url{https://code.mpimet.mpg.de/projects/cdo}). Input data and part of the code used for this study are available through a dedicated \texttt{GITHUB} repository (\url{https://github.com/progettopulvirus/A_spatiotemporal_analysis_of_NO2_concentrations}).

\section{Results and discussion}\label{Sec:results}

\subsection{Model parameters}

\begin{figure}[!ht]
\includegraphics[width=\textwidth]{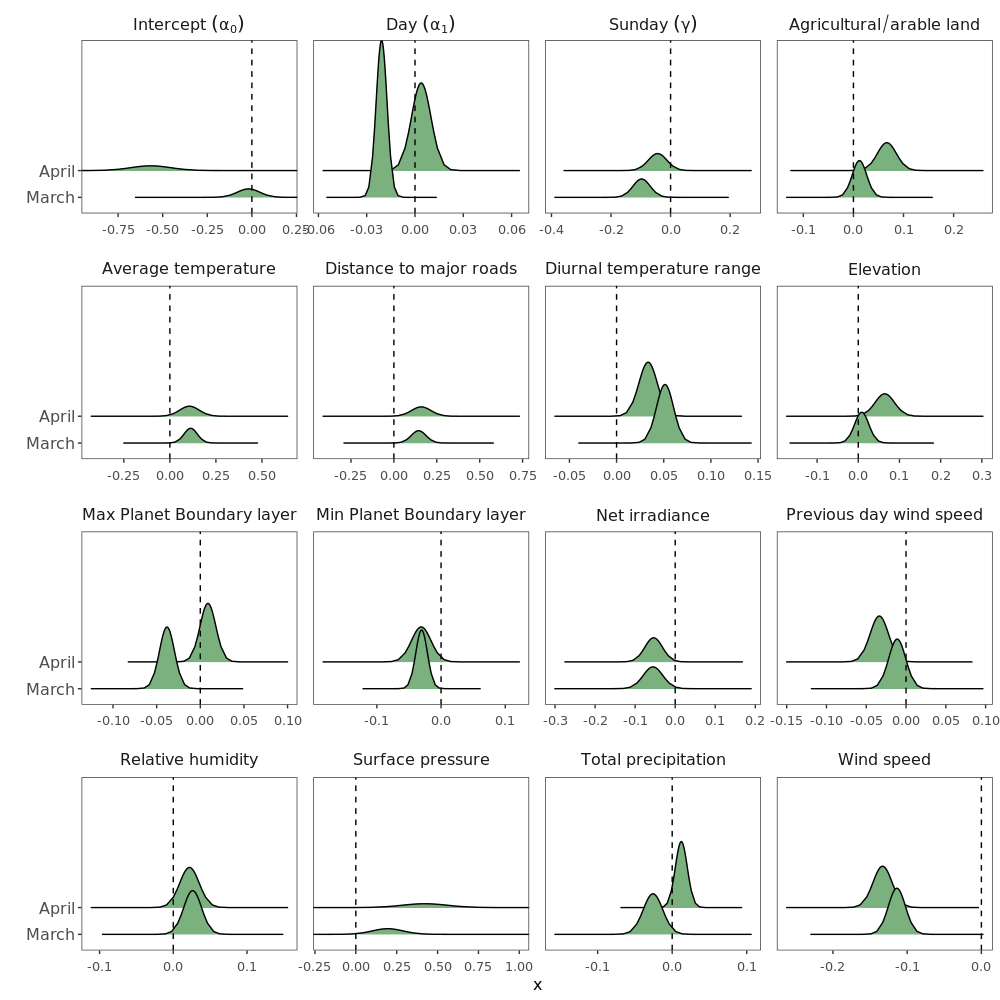}
\caption{Posterior distribution of $\alpha_0$ (Intercept), $\alpha_1$ (Day), $\gamma$ (Sunday) and  of the covariate coefficients $\boldsymbol{\beta}_x$ and $\boldsymbol{\beta}_z$. When the bulk of the posterior distribution lies far from zero it can be concluded that the corresponding parameter is significantly different from zero.}\label{fig:ggridge}
\end{figure}

Figure~\ref{fig:ggridge} shows the monthly posterior distributions for the fixed effects' coefficients ($\alpha_0$, $\alpha_1$, $\gamma$, $\bm{\beta}_z$ and $\bm{\beta}_x$). Generally speaking, we observe that a significant effect, invariant in sign across the months of March and April, characterizes most of the selected regressors. The shape of the posterior distributions is rather stable in the models for the two months, but some exceptions are apparent. The distribution of the linear trend coefficient $\alpha_1$  is narrower in March than in April, while the opposite happens for the posterior distribution of the intercept $\alpha_0$. This is in line with the large variability which characterizes the weekly prediction maps of April (see Section~\ref{Sec:maps}). Finally, the surface pressure coefficient exhibits two almost flat posterior distributions, but this result is not of easy interpretation. 

A significant negative linear trend coefficient $\alpha_1$ was found in March, which suggests a decreasing trend for the daily differences of the log-transformed NO$_{2}$ concentrations in the month when the lockdown restrictions occurred. A significant negative effect was also found for the Sunday coefficient $\gamma$. We can interpret this result saying that a weekly cycle persists even when we consider the difference of the log-transformed NO$_{2}$ concentrations. Surprisingly, neither the linear trend nor the Sunday dummy variable coefficients are significant in April.

With regards to the meteorological parameters, we distinguish those with a positive significant effect (the diurnal temperature range, the average temperature at 2 meters, the relative humidity and surface pressure) and those with a negative significant effect (the min and max planet boundary layer, the net irradiance, the wind speed and the total precipitation). 
Conversely, all the spatial regressors (elevation, \% of agricultural/arable land and distance to major roads) show a positive posterior mean. This could suggest that far from the urbanized centers and the road network the level of the NO$_{2}$ daily concentrations in 2020 tends to be equal or greater than the ones in 2019.

Table~\ref{tab:hyper} provides the posterior summary statistics for the remaining model parameters. We observe that the posterior mean of the AR(1) autocorrelation coefficient $a$, the spatial range $\rho$ and the standard deviation $\sigma_v$ have a larger posterior mean in April than in March.

\begin{table}[!htbp] \centering 
\begin{tabular}{@{\extracolsep{0.5pt}} lccccc} 
\toprule
\phantom{abc} & $a$ & $\rho$ & $\sigma_v$  & $\sigma_{\epsilon}$ & $\sigma_{\omega}$ \\ 
\midrule 
March & 0.64 (0.023) & 74 (4.0) & 0.16 (0.011) & 0.21 (0.004) & 0.37 (0.013) \\ 
April & 0.80 (0.021) & 97 (5.5) & 0.22 (0.015) & 0.22 (0.003) & 0.42 (0.021) \\ 
\bottomrule
\end{tabular} 
  \caption{Posterior mean and standard deviation (in parentheses) of the  parameters in the models for March and April. $a$: AR(1) coefficient; $\rho$ and $\sigma_{\omega}$: range (in km) and standard deviation of the Mat\'ern spatial covariance function; $\sigma_v$: standard deviation of the small scale spatial random effect; $\sigma_{\epsilon}$: standard deviation of the Gaussian measurement error.}
  \label{tab:hyper} 
\end{table} 

\subsection{Model validation}

\begin{figure}[!h]
    \subfloat[Training stage - March]
    {\includegraphics[width=0.45\textwidth]{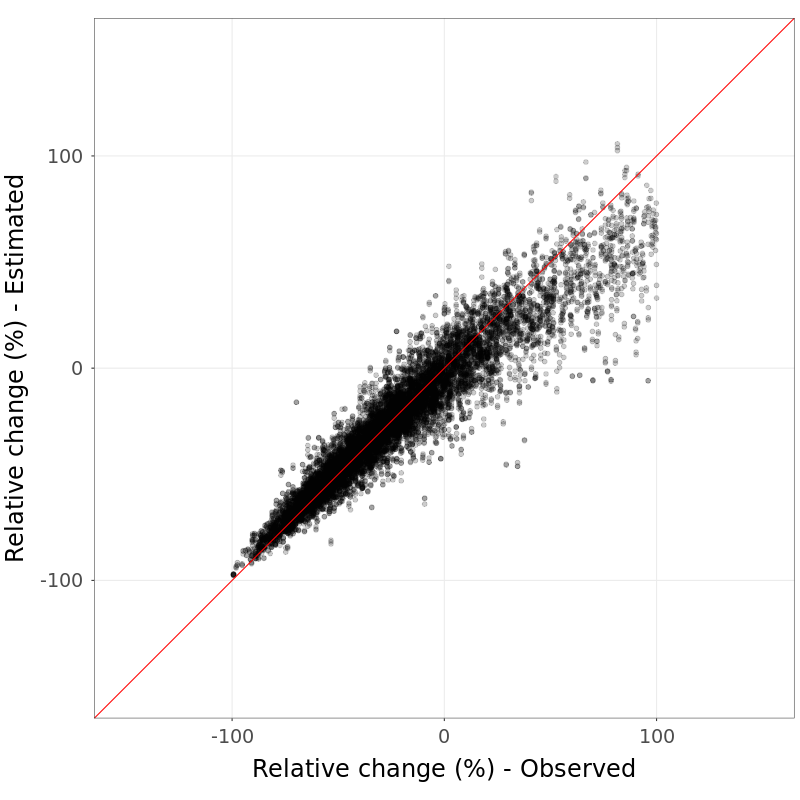}}%
    \qquad
    \subfloat[Training stage - April]{\includegraphics[width=0.45\textwidth]{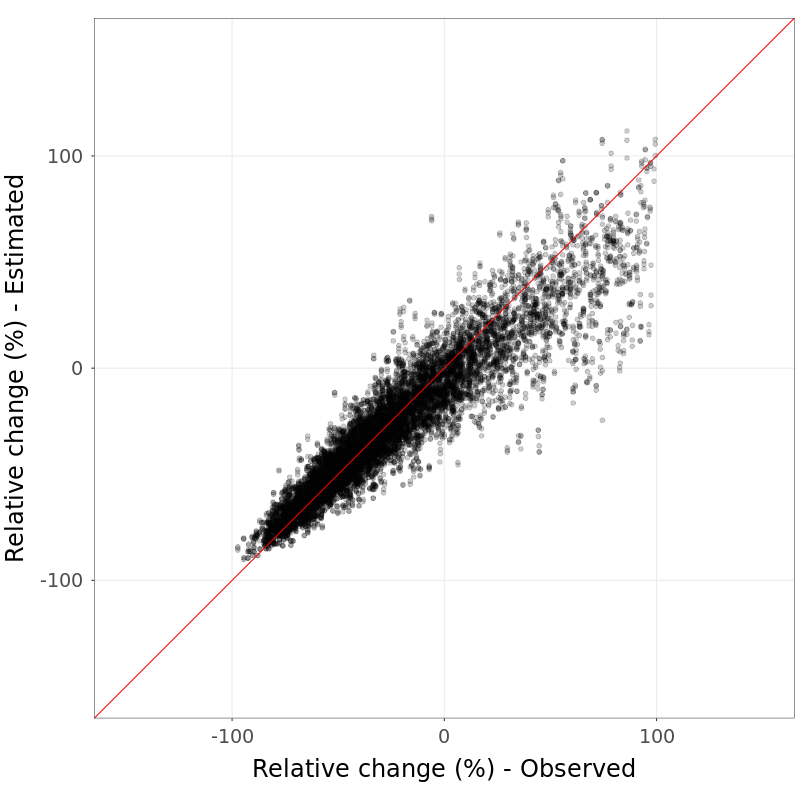}}%
    \qquad
    \subfloat[Validation stage - March ]{\includegraphics[width=0.45\textwidth]{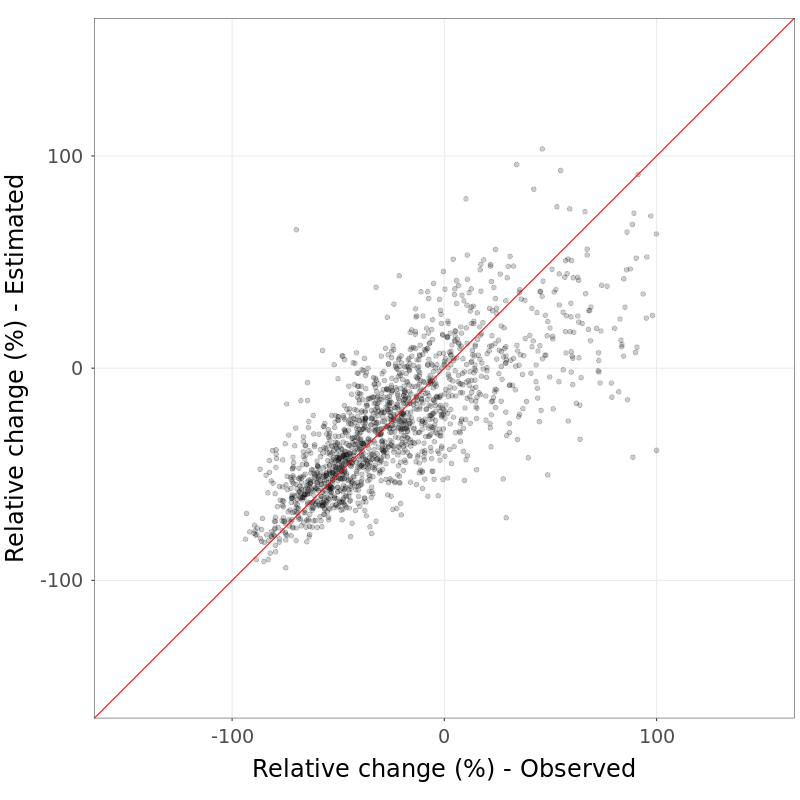}}%
    \qquad
    \subfloat[Validation stage - April]{\includegraphics[width=0.45\textwidth]{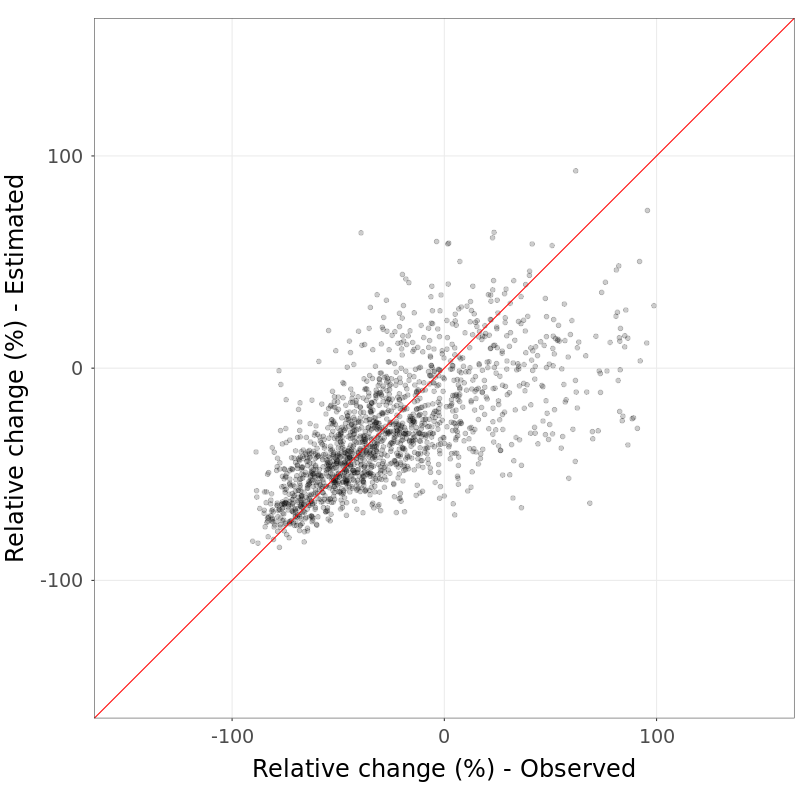}}%
    \caption{Agreement between modelled and measured $\text{NO}_{2}$ relative change ($\tilde\Delta(t,s)\%$) in the training (top) and validation stage (bottom). The solid red line is the 1:1 line as a reference.}%
    \label{fig:scatterplot}%
\end{figure}

For validation purposes, we stratified the input monitoring sites according to their type classification (urban, suburban and rural stations). Within each of these groups, we sampled 10\% of the stations in order to define a validation dataset. The remainder of the stations (training dataset) was used to fit the model and the fitted model was used to predict the daily differences of the log-transformed NO$_{2}$ concentrations (see Equation~\eqref{eq:modelEQ1}) on the validation dataset. Both the sampling and the estimation process were repeated three times for each month. Comparing predicted and observed values allows to investigate how the model performes on unseen data, specifically if the model  generalizes well or suffers from overfitting/underfitting.

Plots and summary statistics are shown as percentage relative changes $\tilde \Delta(t,s) \%$ (see Equation~\eqref{eq:exptrans}) for ease of interpretation of the validation results.
The scatterplots in Figure~\ref{fig:scatterplot} shows the distribution of the fitted values versus the observed values. The points spread uniformly along the diagonal line, showing good agreement between observed and modelled data both in the training and validation stage. As expected, modelled relative change for the validation stage shows greater variability than in the training stage.  Also the Pearson correlation coefficients support the good model performance with a value of 0.9 for the training stage and of $\sim0.7$ for the validation stage. Finally, the Root Mean Squared Error is $\sim10\%$ and $\sim20\%$ for the training and validation stage, respectively.

\subsection{NO$_{2}$ 2019/2020 relative change maps} \label{Sec:maps}

Using the predictors described in Section~\ref{Sec:regressors} as individual raster files, the procedure described in Section \ref{Sec:postsummary} was implemented to obtain maps of the relative change of NO$_2$ concentrations between 2019 and 2020, separately for the weeks of March and April.  

\begin{figure}
    \subfloat[March - First week]
    {\includegraphics[width=0.45\textwidth]{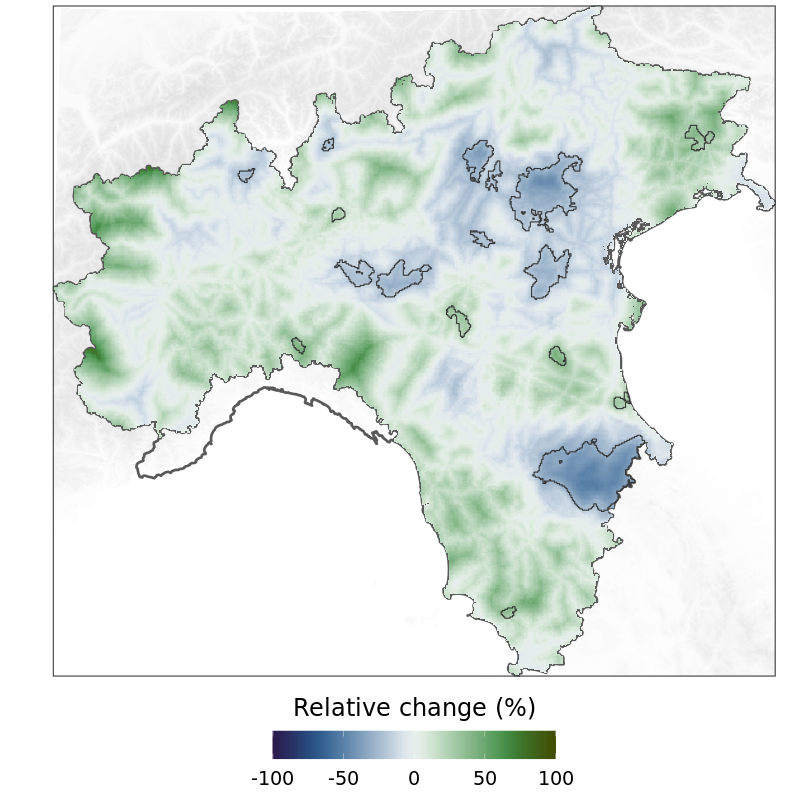}}%
    \qquad
    \subfloat[March - Second week]{\includegraphics[width=0.45\textwidth]{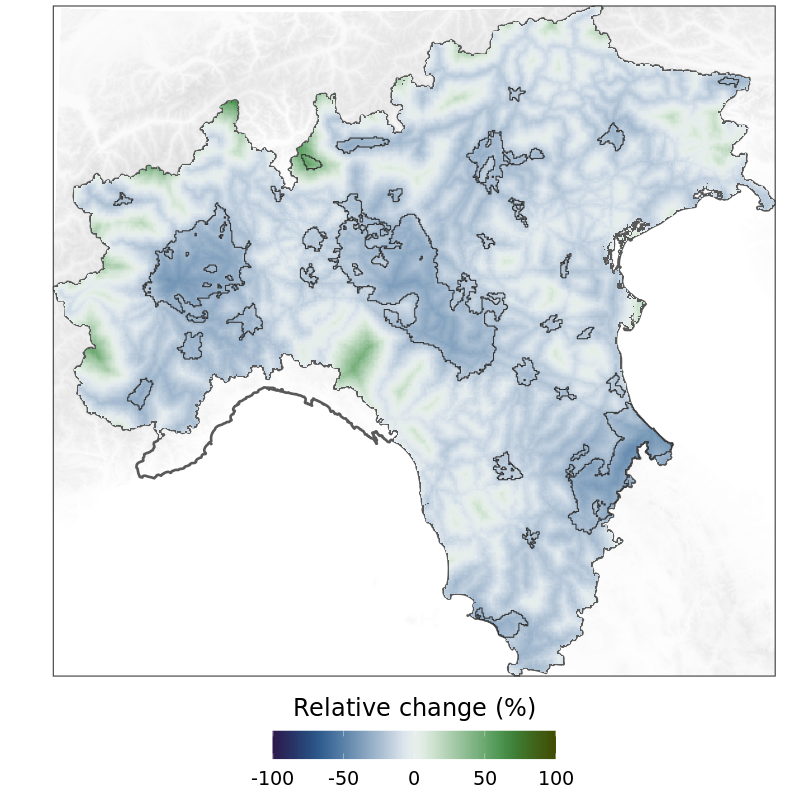}}%
    \qquad
    \subfloat[March - Third week ]{\includegraphics[width=0.45\textwidth]{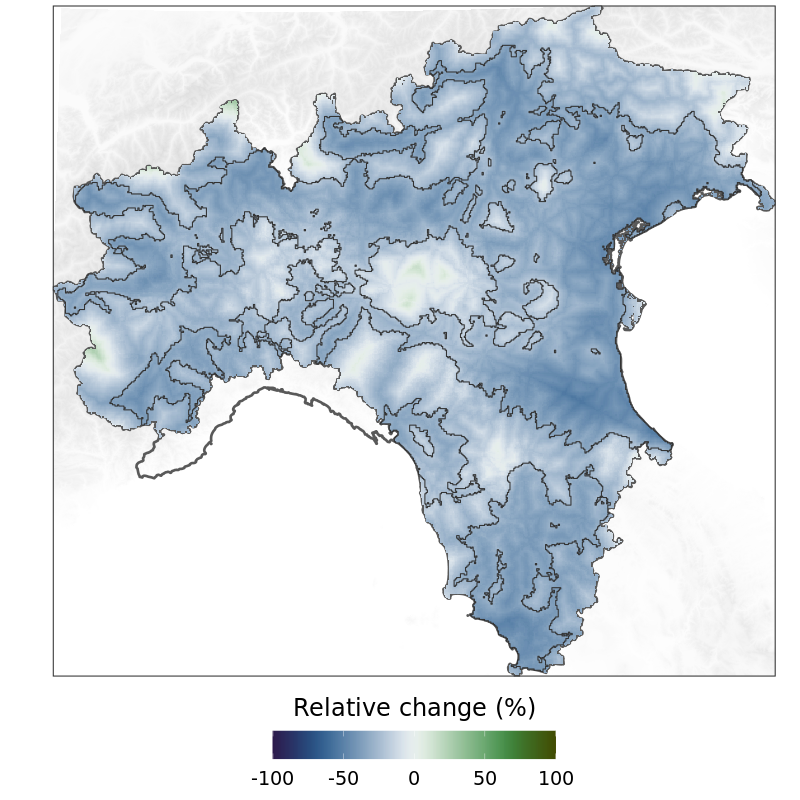}}%
    \qquad
    \subfloat[March - Fourth week]{\includegraphics[width=0.45\textwidth]{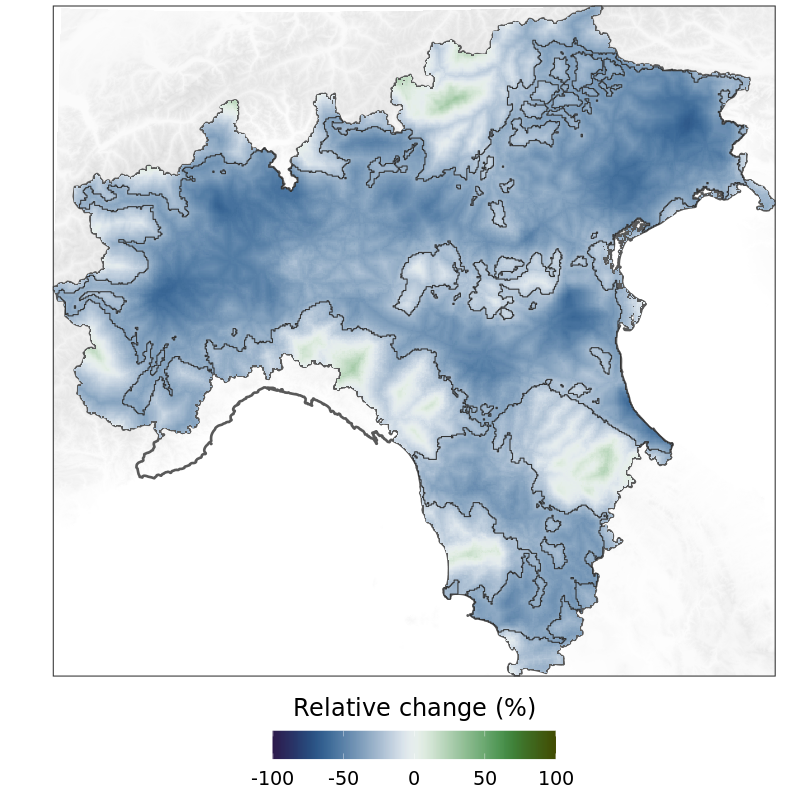}}%
    \caption{Weekly maps of the NO$_{2}$ 2019/2020 relative change (\%)  for the working days (Monday - Saturday). Contour lines mark those areas where the relative change is statistically significant at the significance level of 0.05.}%
    \label{fig:march_maps}%
\end{figure}

\begin{figure}
    \subfloat[April - First week]
    {\includegraphics[width=0.45\textwidth,height=0.3\textheight,keepaspectratio]{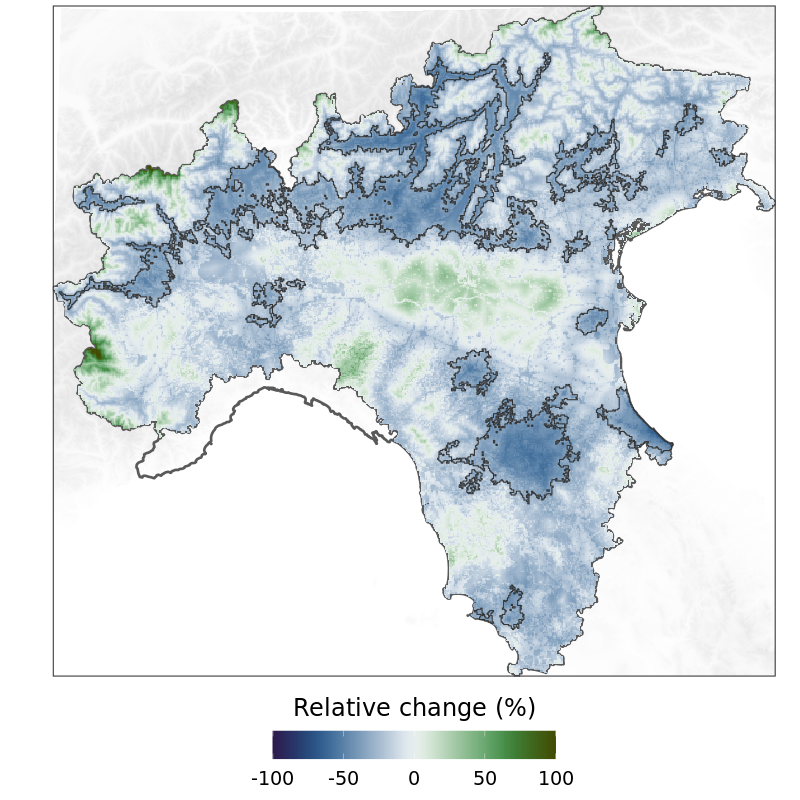}}%
    \qquad
    \subfloat[April - Second week]{\includegraphics[width=0.45\textwidth,height=0.3\textheight,keepaspectratio]{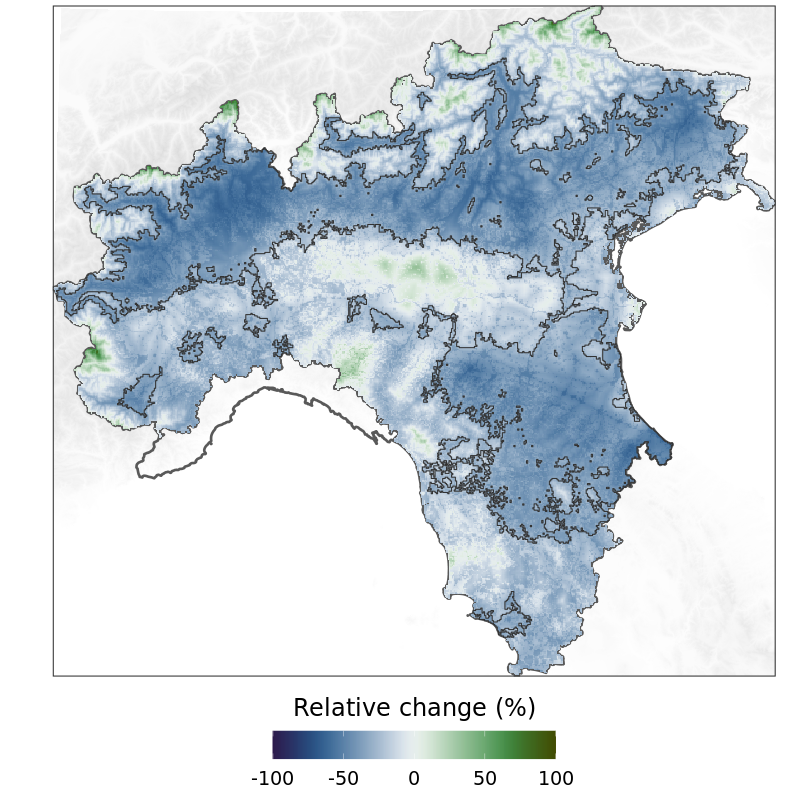}}%
    \qquad
    \subfloat[April - Third week]{\includegraphics[width=0.45\textwidth,height=0.3\textheight,keepaspectratio]{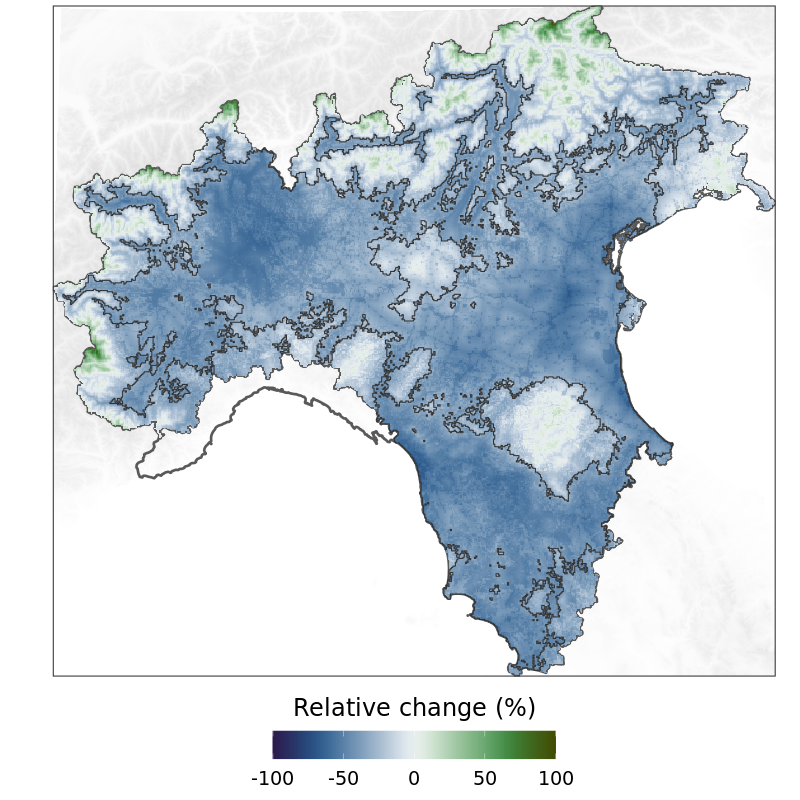}}%
    \qquad
    \subfloat[April - Fourth week]{\includegraphics[width=0.45\textwidth,height=0.3\textheight,keepaspectratio]{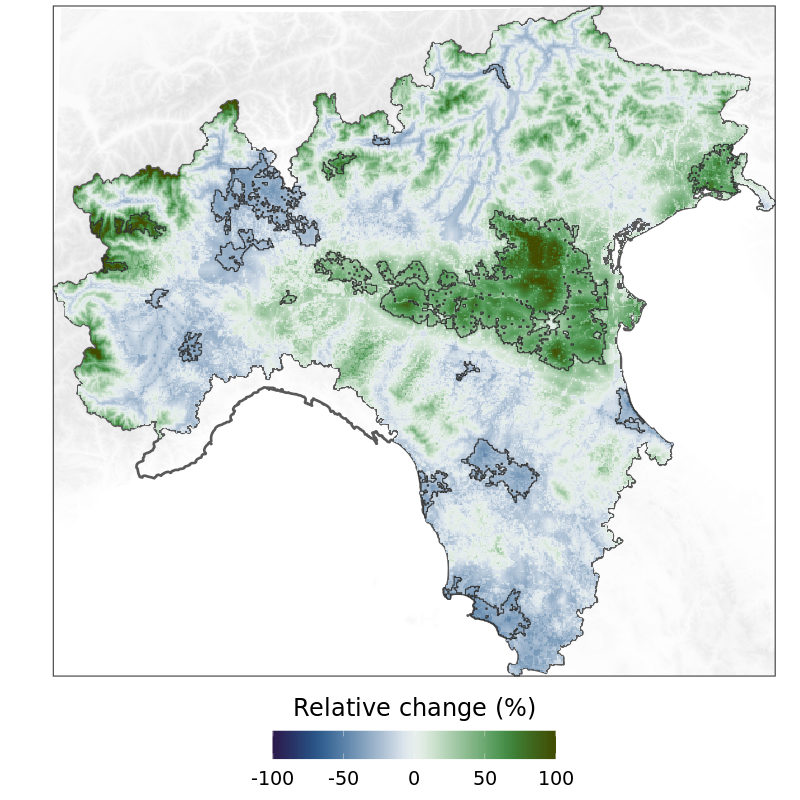}}%
    \qquad
    \subfloat[April - Fifth week]{\includegraphics[width=0.45\textwidth,height=0.3\textheight,keepaspectratio]{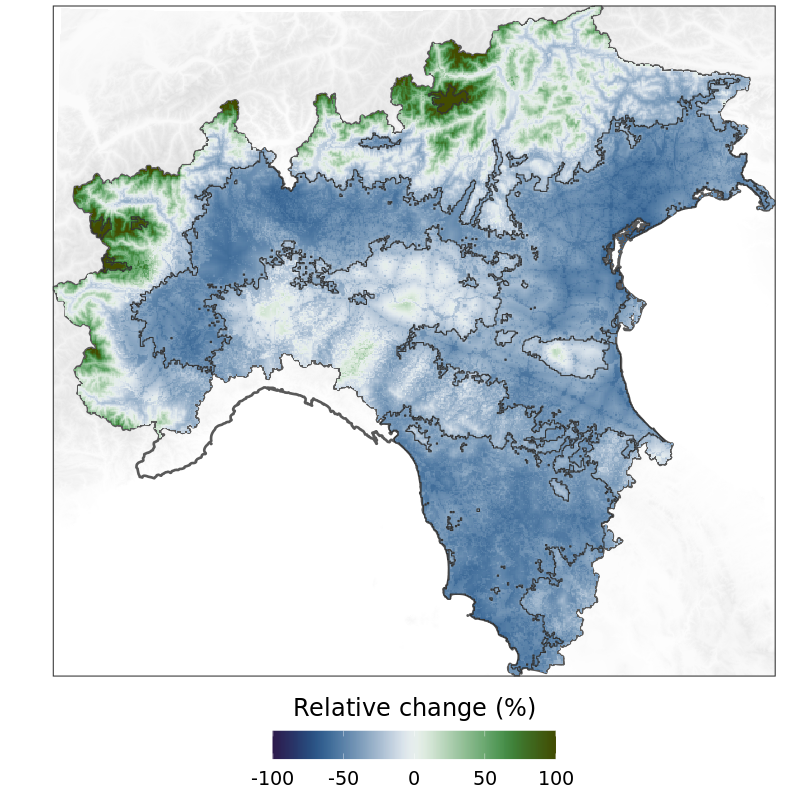}}%
    \qquad
    \hbox to 67.5mm{}
    \caption{Weekly maps of the NO$_{2}$ 2019/2020 relative change (\%) for the working days (Monday - Saturday). Contour lines mark those areas where the relative change is statistically significant at the significance level of 0.05.}%
    \label{fig:april_maps}%
\end{figure}

The maps in Figure~\ref{fig:march_maps} and \ref{fig:april_maps} show the spatial pattern of the 2019/2020 relative change of NO$_{2}$ concentration in the north/centre of Italy for March and April, respectively. These maps refer to the weekly averaged estimates obtained from working days (Monday - Saturday), while the weekly Sunday estimates are available in the Annex (see Section \ref{Sec:annex}). This distinction between working days and Sunday maps reflects our choice of including a dummy regressor for the Sunday effect in the model and its statistical significance observed in March. For visualization purposes, we limited the data range between -100\% and +100\% and used a scientifically derived color map \citep{Crameri2020}. Furthermore, the islands from the Tuscan Archipelago are not displayed and Liguria region, where no daily NO$_{2}$ concentrations were available, is represented with a white background. 

It is worthwhile to stress that the maps presented in this section must be intended as meteorology-normalized maps. This means that they were generated assuming that, in each cell of the raster grid, the daily meteorological conditions are exactly the same in 2019 and 2020. Mathematically, this assumption is equivalent to setting to zero all the meteorological terms in our spatio-temporal model (see Section~\ref{Sec:postsummary}). 

\begin{figure}
    \subfloat[March]
    {\includegraphics[width=0.45\textwidth]{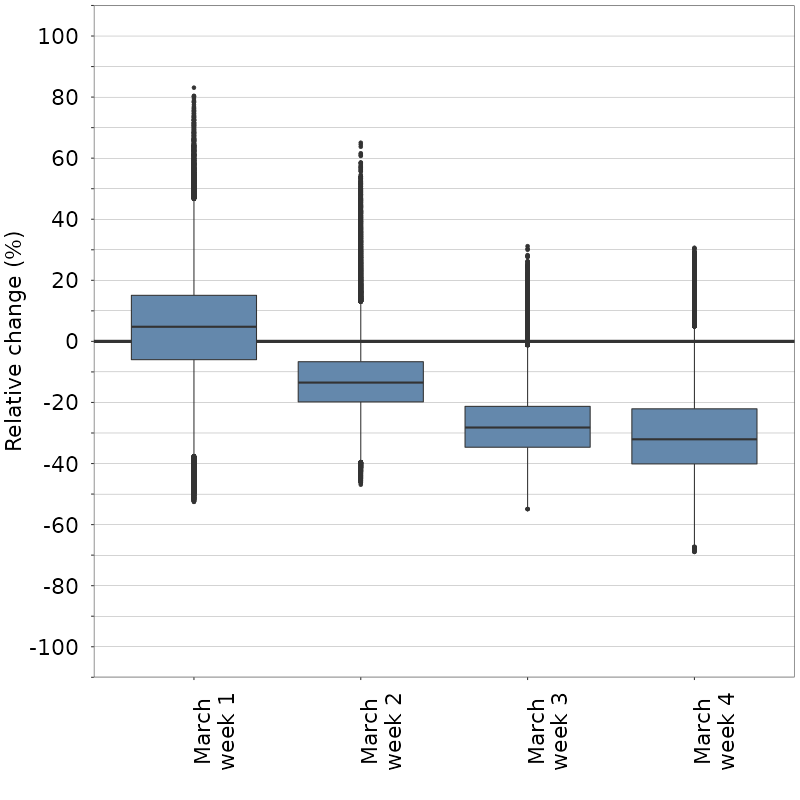}}%
    \qquad
    \subfloat[April]{\includegraphics[width=0.45\textwidth]{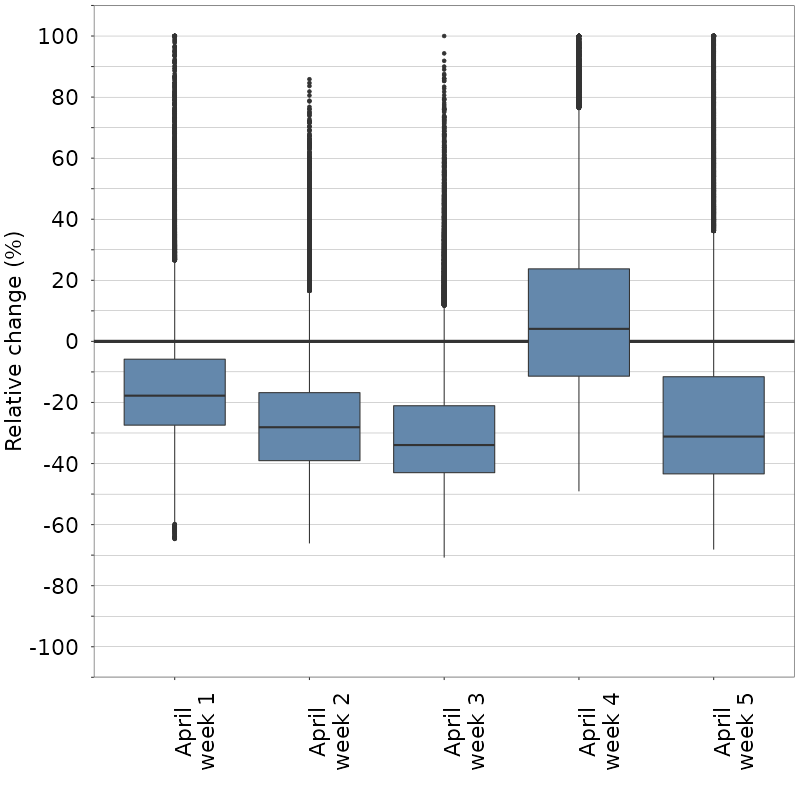}}%
    \caption{Distribution of the NO$_{2}$ 2019/2020 relative change (\%) for the working days (Monday - Saturday) maps.}%
    \label{fig:boxplot_maps}%
\end{figure}

\begin{figure}
\includegraphics[width=\textwidth]{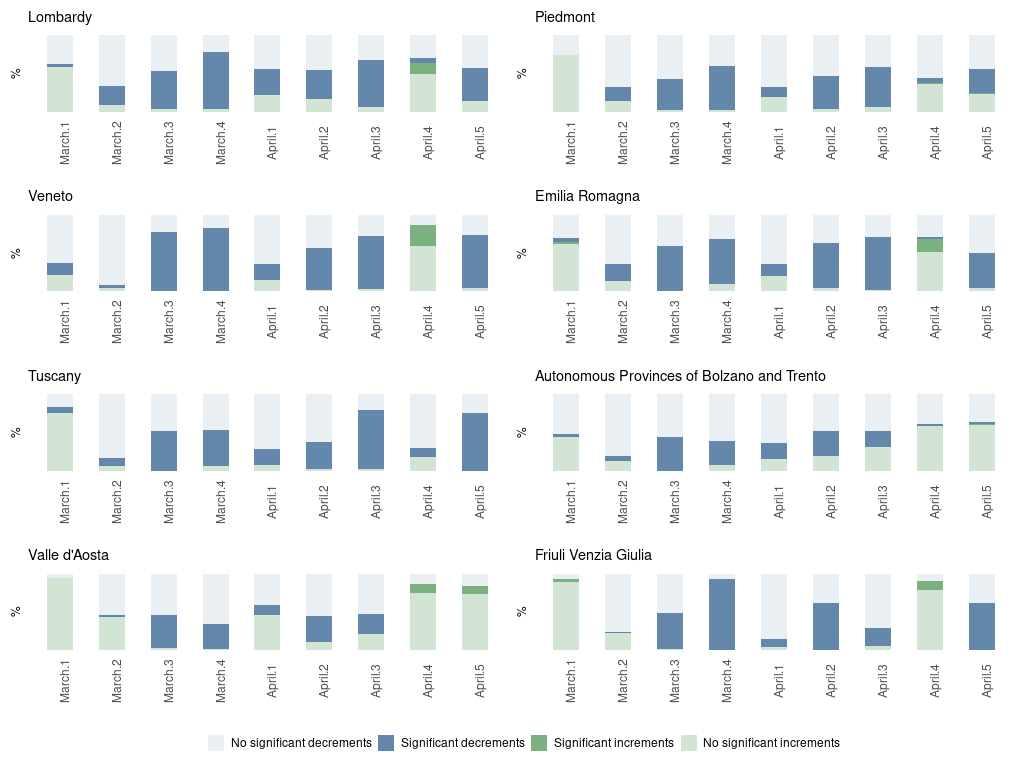}
\caption{Percentage of cells in the prediction maps for working days (Monday - Saturday) with positive (green color scale) and negative (blue color scale) 2019/2020 relative changes. Darker colors are used for significant variations.}\label{fig:barplot_mappe_regionali}
\end{figure}

Figure~\ref{fig:march_maps} and \ref{fig:april_maps} reveal a substantial decrease in NO$_{2}$ concentrations during March and April 2020 as compared to 2019. A statistically significant reduction persists during the third and fourth working weeks of March across the whole study area: we quantified the interquartile range of the corresponding relative changes distribution to be between -40\% and -20\%, as shown in the boxplot in the left panel of Fig.~\ref{fig:boxplot_maps}. Notably, the third and fourth week of March correspond to the stringent phase of the COVID-19 lockdown. The barplots of Figure~\ref{fig:barplot_mappe_regionali} indicate that Lombardy, Veneto, Friuli Venezia Giulia and Emilia Romagna are the regions where significant reductions in the NO$_{2}$ concentrations mostly occurred. Conversely, Tuscany region in the centre and the mountainous regions of Valle d'Aosta and the Autonomus Provinces of Bolzano and Trento are those where most of the estimated changes have no statistical significance.

In April the predicted surfaces show more spatial variability than in March and all the weekly distributions of the relative changes (right panel of Fig.~\ref{fig:boxplot_maps}) exhibit positive increments up to 100\%. These results suggest that the NO$_{2}$ concentrations began to recover in April.  Surprisingly, the fourth week of April is overwhelmed by a significant increment in the agricultural/arable area of the Po Valley (median increment around 10\%), while a spatially homogeneous decrease, like the one observed in the second half of March, occurs only during the third week of the month (interquartile range between -40\% and -20\%). Note that during most of April 2020 the lockdown restrictions were the same of the third and fourth week of March. Looking at the maps of Figure~\ref{fig:april_maps}, it is apparent that the urbanized belt of the Po valley shows a persistent significant negative change during almost all the weeks of April. This is confirmed by the barplots of Figure~\ref{fig:barplot_mappe_regionali}. Despite this variegated situation, the weekly distributions of the relative changes still are dominated by negative values with median values around -30\%, with the exception of the fourth week. 

Generally speaking, the estimates from the Sunday maps (see Section~\ref{Sec:annex}) are consistent with those obtained for working days. The most pronounced decline of the NO$_{2}$ concentrations occurs during the last two Sundays of March: we find a -40\% median relative change in 2020 as compared to the same period in 2019. The same drop in the NO$_{2}$ concentrations is quantified for the second week of April (when Easter 2020 occured).

\newpage

\hfill

\hfill

\clearpage

\newpage

\hfill

\hfill

\clearpage

\section{Conclusions}\label{Sec:Conclusions}

Capturing the spatio-temporal variation in pollutant concentrations is of keen interest for the air pollution management's community. One of the reasons is that the effectiveness of environmental policies or interventions, or even exceptional events as the COVID-19 lockdown, must be evaluated across the whole study domain, in order to understand if the change is homogeneous in space or peculiar spatial patterns occur. This analysis is made even more difficult by the fact that air pollution dynamics strongly depends on weather conditions; thus, it is necessary to disentangle the effect of the adopted intervention by controlling for the known confounding factors. Chemical transport model can be used for this purpose. They can assess, with a very detailed spatial and temporal resolution, variations due to changes in the emission burden arising from interventions, like the implementation of air quality plans. However, some drawbacks exist: CTMs need very detailed input data (e.g., emission inventories, meteorology data), are complex to be run and rely on advanced IT infrastructure for their simulations. More importantly, they do not provide an estimate of the uncertainty of the final predictions which are also strongly dependent on the initial and boundary conditions. Moving to statistical models, it is straightforward to deal with uncertainty and spatio-temporal correlation structures, while keeping the computational costs at a reasonable level (especially if computationally efficient estimation methods, like the INLA-SPDE approach, are used).

In this paper we propose the use of a statistical Bayesian spatio-temporal model as a novel tool for assessing - in time but more importantly in space - the effectiveness of air quality policy interventions. As a case study, we focused on how the COVID-19 lockdown affected air pollution levels in the north/centre of Italy, by means of the 2019/2020 NO$_{2}$ concentrations relative change, the main output of our modeling strategy. Even if our proposal does not explicitly simulate physical and chemical reactions in the atmosphere (like CTMs do), it has two key advantages: it is parsimonious in terms of data needs and, possibly more important, it naturally manages the uncertainty associated to the parameters and map outputs. The proposed method, applied to a real-world experiment, demonstrated its capability to provide a reliable picture of the temporal and spatial patterns of the NO$_{2}$ variations (compared to 2019), while accounting for the effect of meteorology. 

In this study, the spatio-temporal variation of the NO$_{2}$ concentrations in March and April 2020, as compared with the same period in 2019, is illustrated through meteorology-normalized spatially continuous maps at weekly intervals, both for working days (Monday - Saturday) and weekends (Sunday). Our results show that during March and April 2020 the study domain was generally characterized by negative relative changes in the NO$_{2}$ concentrations with median values around -25\%. Such estimate seems to be reasonably consistent with previous findings about NO$_2$ levels during the lockdown in Europe. However, the message from our output maps is richer than what a simple number can describe. First, a visual inspection of the maps shows that statistically significant reductions mainly occurred in the urban areas of the Po valley and Tuscany, while not significant variations persisted over the mountainous regions of Valle d'Aosta and the Autonoumous Provinces of Bolzano and Trento. Second, and more interestingly, our weekly analysis of NO$_2$ highlight that in March 2020 such reduction was synchronous with the lockdown measures adopted, while in April 2020 the concentrations, as compared to 2019, started to recover in some parts of the investigated area. To the authors of this study, this result comes not unexpected as it reflects the behaviour of the input data seen in the parallel plot of Figure~\ref{fig:serie_osservate}. At the same time, it comes as a surprise that none of the statistical studies we examined documented an increase in the NO$_{2}$ concentrations in April in the studied Italian domain. However, we are aware that previous findings are based on different models, data, and methodologies, so our results are not directly comparable, especially if we consider that all these studies do not provide continuous maps as an output.

It is worthwhile to observe that our modeling approach does not allow us to draw causal inference conclusions. However, given that the maps account for the weather effect, we can conclude that the reductions we observe across regions and weeks can be attributed to a factor different from the meteorology. Given that they occur in the same period of the restrictive measures, it is likely that the COVID-19 lockdown had an effect in the reduction of air pollution. Nevertheless, we can not exclude that other factors, not considered in the model, could have ad an active role in the dynamics of NO$_2$.

The model described in this paper takes inspiration from the model introduced in \cite{FIORAVANTI2021} for the spatio-temporal assessment of the daily log-transformed PM$_{10}$ concentrations in Italy. As here our objective was to describe the relative changes across the two considered years (2019 vs 2020), in this study the target variable is the daily differences of the log-transformed NO$_{2}$ concentrations. As far as we know this modeling strategy has never been used before in the context of spatio-temporal models for intervention analysis. Other approaches could have been adopted, still in the context of spatio-temporal modeling. For example, the same problem could have also been tackled by, first, jointly modeling the 2019 and 2020 NO$_{2}$ concentrations and then computing the differences between the daily gridded predictions. However, this alternative solution has two important shortcomings: 1) it is not consistent with the AR(1) assumption, as our observations are not consecutive in time but are one year apart (March-April 2019 and 2020); 2) it is more time consuming as the size of the input dataset would be as twice as the size of the input dataset when daily differences are considered. Furthermore, the use of daily differences alleviates multicollinearity among regressors and allows to include more parameters in the regression equation. For this reason we believe that the spatio-temporal model we propose represents a valid solution to analyze pollutant concentrations measured by a network of monitoring stations before and after a given event of interest (e.g., the lockdown intervention). Interestingly, our model allows a straightforward spatial assessment of NO$_{2}$ variations through high-resolution estimates, since the posterior predictive distributions of NO$_{2}$ differences can virtually be derived for every pixel within the study domain. 
This is of keen interest for policy makers involved in intervention analysis or epidemiologist assessing population exposure change and gradients, since the assessment at urban hot spot, the between-city variability as well as the comparison among rural and urban context can be achieved at the finest resolution, with a reasonable computational effort. In this regard, our approach could be usefully replicated whenever the effects of an intervention aimed to reduce the pollutant emissions at local or regional scale have to be assessed. Last but not least, the proposed model is simple and can be easily extended to a larger spatial domain. This means that it could be also applied to other pollutants or, more generally, spatio-temporal phenomenon which are continuous in space and for which it is interesting to evaluate the change in consecutive years.

\section{Appendix}\label{Sec:annex}

\newpage

\begin{figure}
    \subfloat[March - First week]
    {\includegraphics[width=0.45\textwidth,height=0.3\textheight,keepaspectratio]{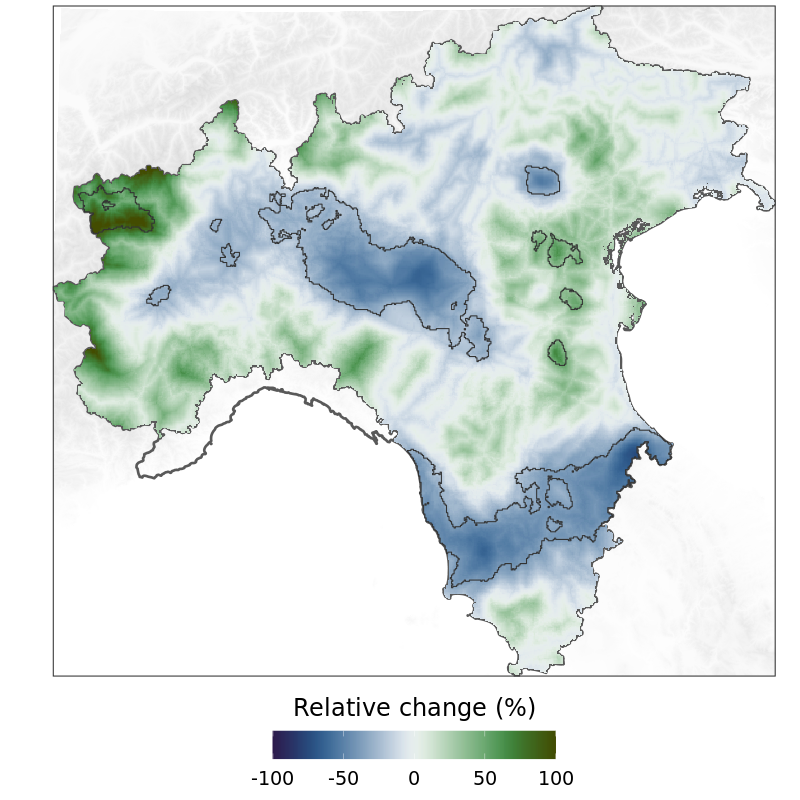}}%
    \qquad
    \subfloat[March - Second week]{\includegraphics[width=0.45\textwidth,height=0.3\textheight,keepaspectratio]{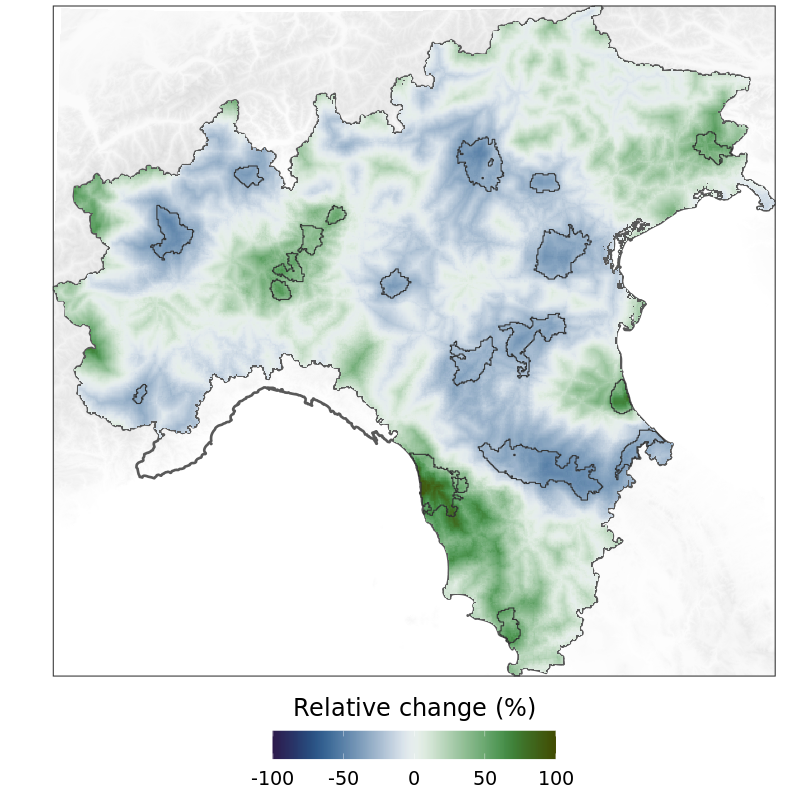}}%
    \qquad
    \subfloat[March - Third week ]{\includegraphics[width=0.45\textwidth,height=0.3\textheight,keepaspectratio]{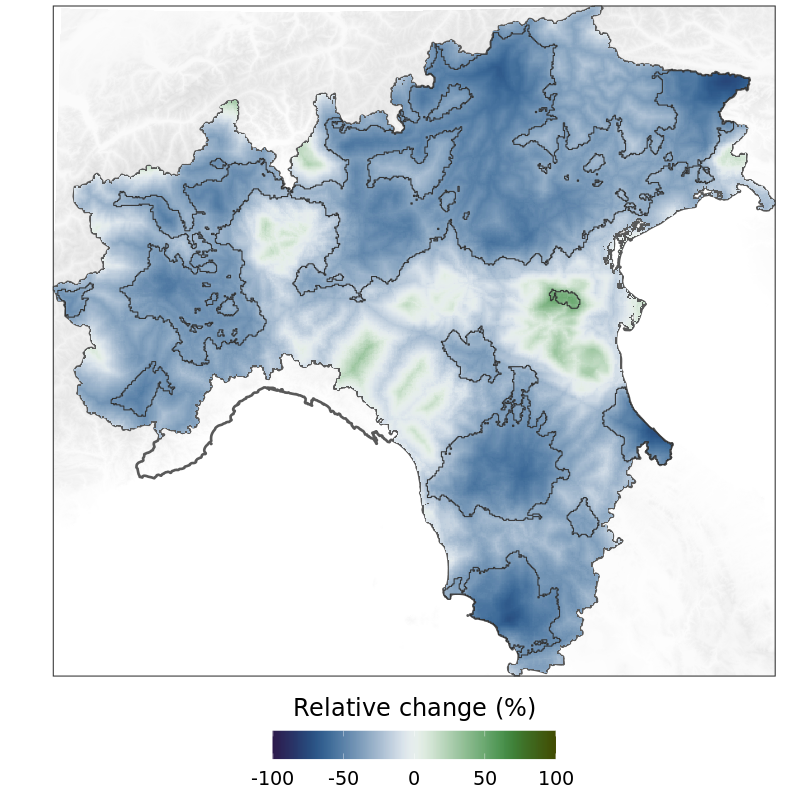}}%
    \qquad
    \subfloat[March - Fourth week]{\includegraphics[width=0.45\textwidth,height=0.3\textheight,keepaspectratio]{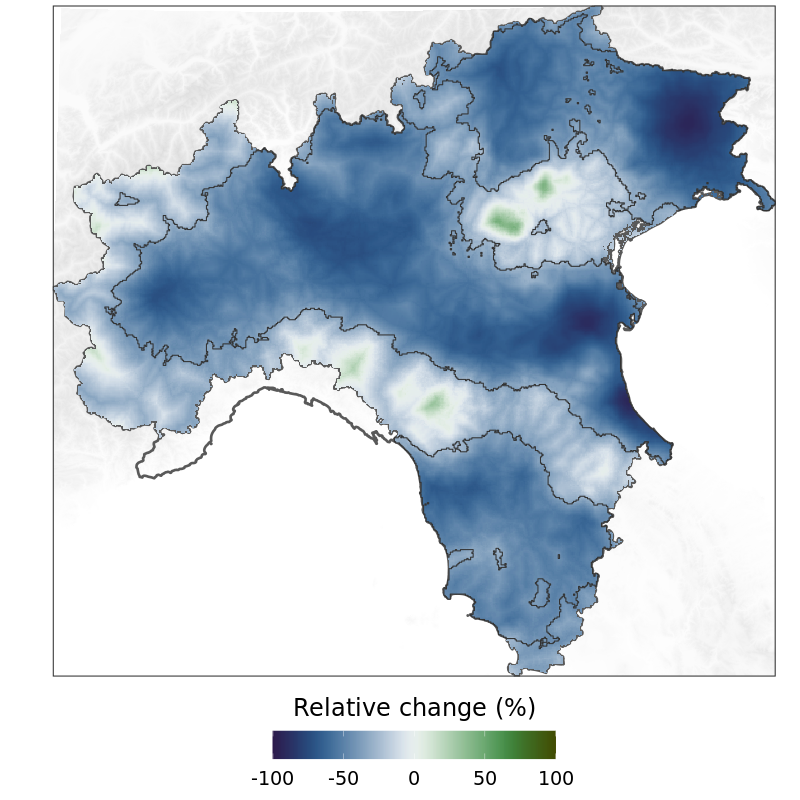}}%
   \qquad
    \subfloat[March - Fifth week]{\includegraphics[width=0.45\textwidth,height=0.3\textheight,keepaspectratio]{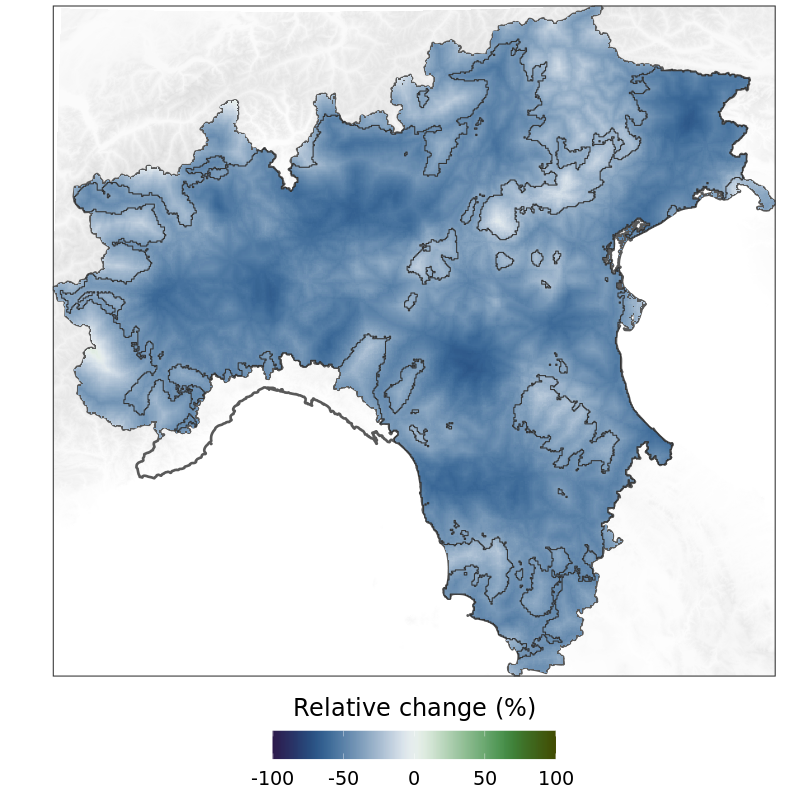}}%
    \qquad
    \hbox to 67.5mm{}
    \caption{Weekly maps of the NO$_{2}$ 2019/2020 relative change (Sunday). Contour lines mark those areas where the relative change is statistically significant at the significance level of 0.05.}%
    \label{fig:march_maps_festivi}%
\end{figure}

\begin{figure}
    \subfloat[April - First week]
    {\includegraphics[width=0.45\textwidth]{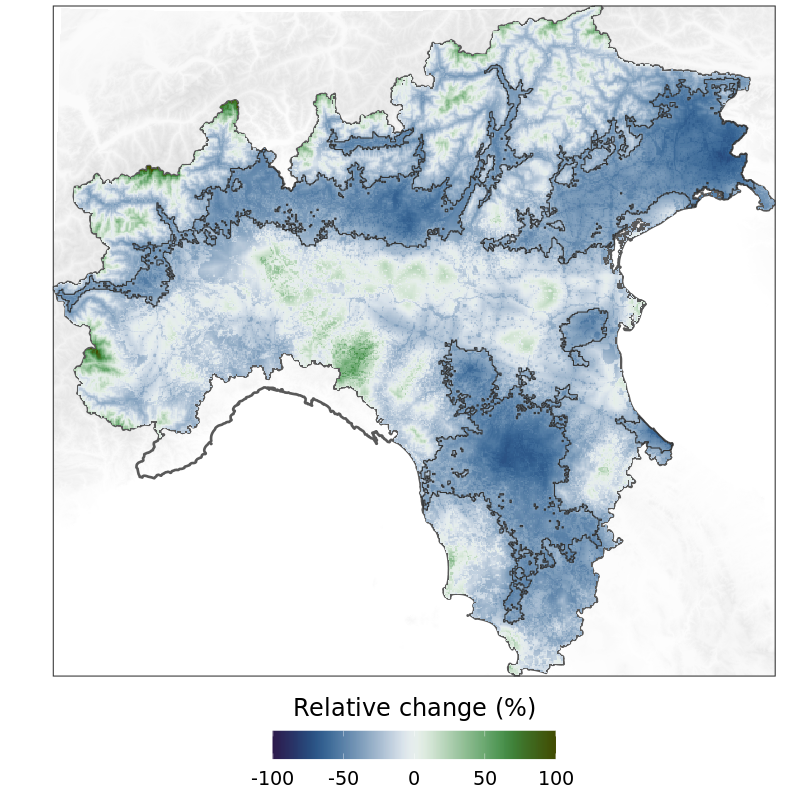}}%
    \qquad
    \subfloat[April - Second week]{\includegraphics[width=0.45\textwidth]{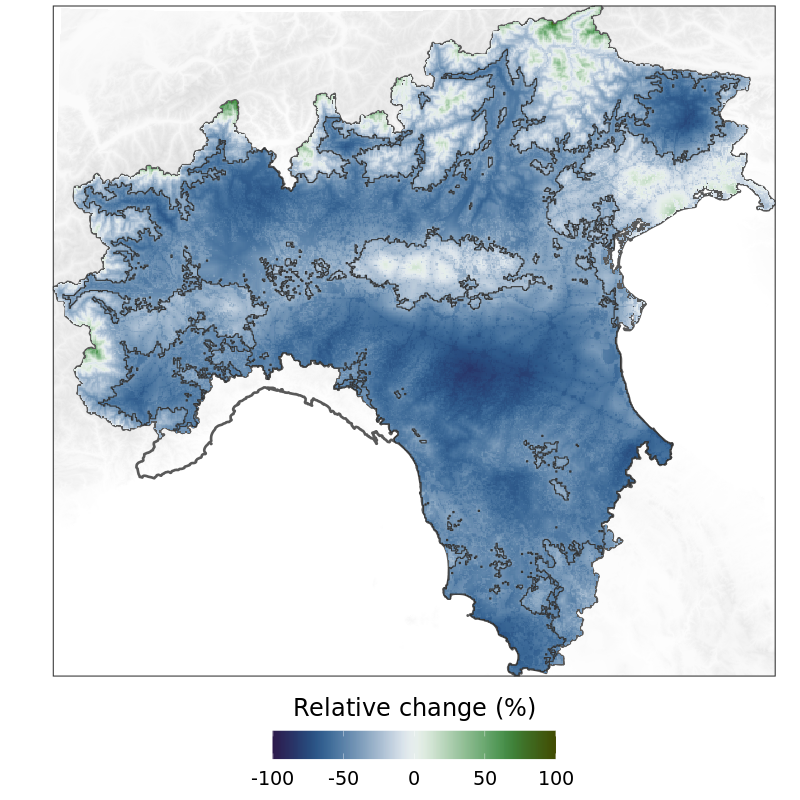}}%
    \qquad
    \subfloat[April - Third week]{\includegraphics[width=0.45\textwidth]{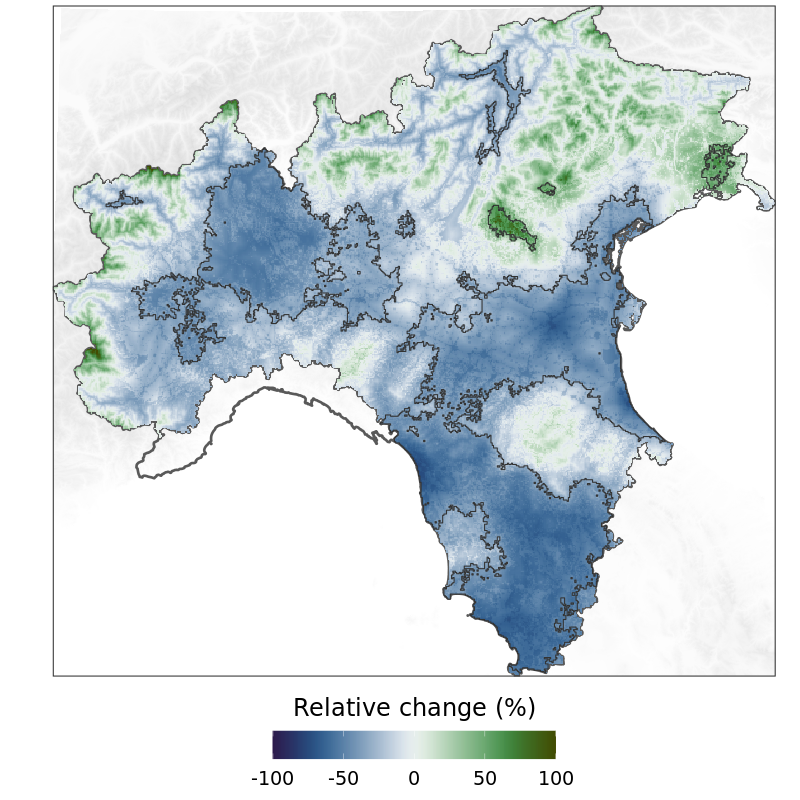}}%
    \qquad
    \subfloat[April - Fourth week]{\includegraphics[width=0.45\textwidth]{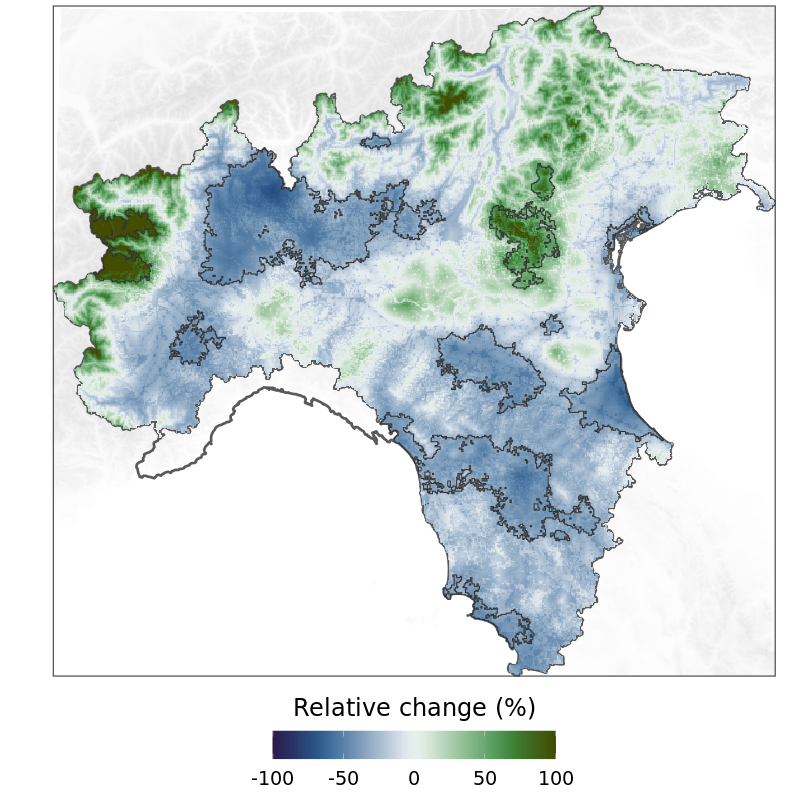}}%
    \caption{Weekly maps of the NO$_{2}$ 2019/2020 relative change (Sunday). Contour lines mark those areas where the relative change is statistically significant at the significance level of 0.05.}%
    \label{fig:april_maps_festivi}%
\end{figure}

\clearpage

\begin{figure}
    \subfloat[March]
    {\includegraphics[width=0.45\textwidth]{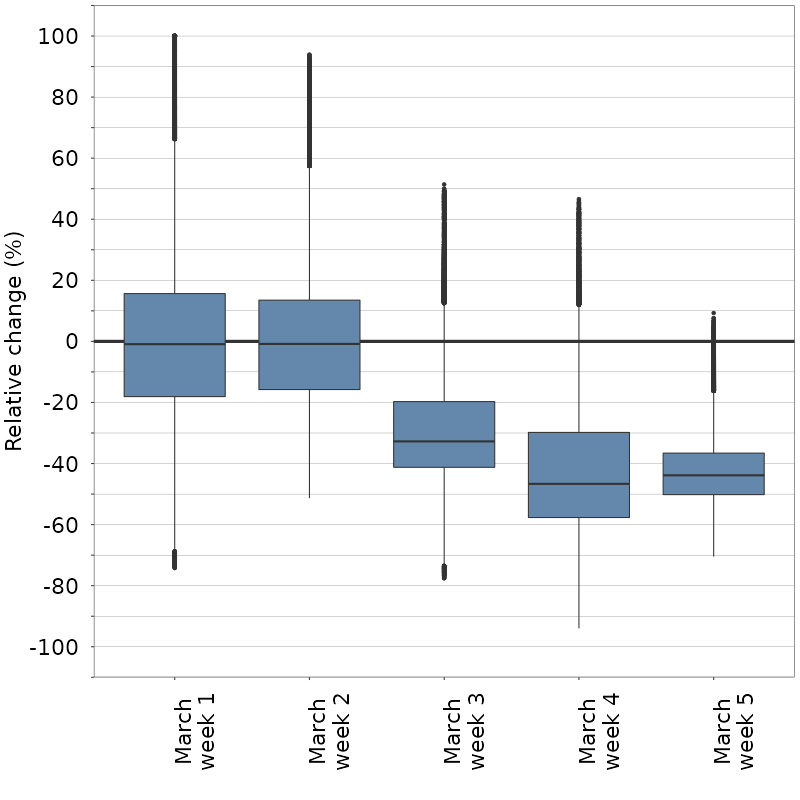}}%
    \qquad
    \subfloat[April]{\includegraphics[width=0.45\textwidth]{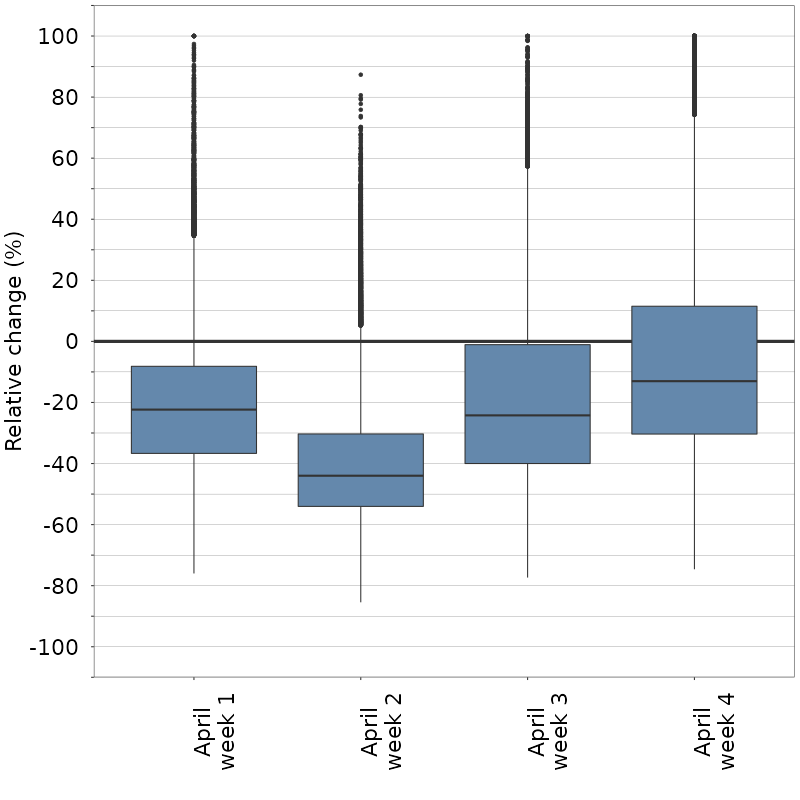}}%
    \caption{Distribution of the estimated NO$_{2}$ 2019/2020 relative change (\%) for the Sunday maps.}%
    \label{fig:boxplot_maps_festivi}%
\end{figure}

\section{Acknowledgments}

This research has been carried out within the framework of the PULVIRUS partnership agreement between ENEA, ISPRA, SNPA and ISS.

We wish to thank for their helpful contribution to the air quality and meteorological data pre-processing Raffaele Morelli, Walter Perconti and Arianna Trentini. In addition, we wish to thank all the Regional Environmental Protection Agencies (ARPA Valle d’Aosta, ARPA Piemonte, ARPA Lombardia, ARPA Veneto, APPA Bolzano, APPA Trento, ARPA Friuli, ARPAE Emilia Romagna, ARPA Toscana) that manage the air quality networks and provide fully validated NO$_{2}$ hourly data.

This research did not receive any specific grant from funding agencies in the public, commercial, or not-for-profit sectors. 
\subsection*{Author contributions}

Guido Fioravanti, Michela Cameletti and Sara Martino conceived the presented model. 
Guido Fioravanti prepared the data with the support of Giorgio Cattani and performed computations. Michela Cameletti took the lead in writing the manuscript. Sara Martino verified the analytical methods. Giorgio Cattani and Enrico Pisoni contributed to the interpretation of the results. All authors discussed the results and contributed to the final version of the manuscript. 

\subsection*{Financial disclosure}

None reported.

\subsection*{Conflict of interest}

The authors declare no potential conflict of interests.


\bibliography{mybibfile}
\end{document}